\documentclass[%
 reprint, 
 superscriptaddress,
nofootinbib,
 amsmath,amssymb,
 aps, 
 prx
]{revtex4-2}

\usepackage{graphicx}
\usepackage{dcolumn}
\usepackage{bm}

\usepackage{hyperref}
\hypersetup{
    colorlinks,
    linkcolor={blue},
    citecolor={blue},
    urlcolor={blue}
}
\usepackage[capitalize]{cleveref}
\Crefname{section}{Sec.}{Secs.}
\Crefname{equation}{Eq.}{Eqs.}
\creflabelformat{equation}{(#2#1#3)}
\crefrangelabelformat{equation}{#3(#1)#4--#5(#2)#6}
\crefrangelabelformat{section}{#3#1#4--#5#2#6}
\crefrangelabelformat{subsection}{#3#1#4--#5#2#6}
\crefrangelabelformat{figure}{#3#1#4--#5#2#6}
\crefrangelabelformat{table}{#3#1#4--#5#2#6}

\usepackage{orcidlink}
\usepackage[separate-uncertainty=true,range-phrase=--,multi-part-units=single,range-units=single]{siunitx}
\usepackage{dsfont,cancel}

\usepackage{etoolbox}
\makeatletter
\patchcmd{\paragraph}{-1em}{-.0001em}{}{ }
\patchcmd{\subparagraph}{-1em}{-.0001em}{}{ }
\makeatother

\newcommand{\kB}{k_{\rm B}}
\newcommand{\action}{{\mathcal{S}}}
\newcommand{\rayleigh}{{\mathcal{R}}}
\newcommand{\Lag}{{\rm L}}
\newcommand{\Eul}{{\rm E}}
\newcommand{\sys}{{\rm s}}
\newcommand{\bath}{{\rm b}}
\newcommand{\any}{{\rm a}}
\newcommand{\ela}{{\rm ela}}
\newcommand{\act}{{\rm act}}
\newcommand{\flu}{{\rm flu}}
\newcommand{\ext}{{\rm ext}}
\newcommand{\com}{C}
\newcommand{\soft}{{\rm soft}}
\newcommand{\shell}{{\rm shell}}
\newcommand{\cell}{{\mathcal{V}}}
\newcommand{\mic}{{\rm mic}}
\newcommand{\helfrich}{{\rm H}}
\newcommand{\mass}{{\mathcal{M}}}
\newcommand{\dbar}{\mathchar'26\mkern-12mu d}
\newcommand{\fr}[1]{\left\|#1\right\|}
\newcommand{\avg}[1]{\left\langle#1\right\rangle}
\newcommand{\mtx}[1]{{\mathsf{#1}}}
\newcommand{\tra}[2][ ]{{#2}^{\operatorname{#1{T}}}}
\DeclareMathOperator{\tr}{tr}
\DeclareMathOperator{\const}{const}

\begin{document}

\newcommand{\SM}[1]{\cite[{#1}]{supp}}
\newcommand{\smNum}{S1}
\newcommand{\smDelta}{S5}
\newcommand{\smParams}{S7}
\newcommand{\eqSurf}{S3}
\newcommand{\movPull}{1}
\newcommand{\movPush}{2}


\title{From Action Principle to Stochastic Thermodynamics of\texorpdfstring{\\}{}Elastic and Viscous Active Matter}%

\author{Valentina Barletta\orcidlink{0009-0003-0814-130X}}
\thanks{These authors contributed equally to this work.}
\affiliation{Cell Biology and Biophysics Unit, European Molecular Biology Laboratory, Meyerhofstraße 1, 69117 Heidelberg, Germany}
\affiliation{Department of Physics, University of Turin, Via Verdi 8, 10124 Turin, Italy}
\author{Michele Lupini\orcidlink{0009-0000-6438-7978}}
\thanks{These authors contributed equally to this work.}
\affiliation{Cell Biology and Biophysics Unit, European Molecular Biology Laboratory, Meyerhofstraße 1, 69117 Heidelberg, Germany}
\author{Ian D. Estabrook\orcidlink{0000-0002-3241-6890}}%
\affiliation{Cell Biology and Biophysics Unit, European Molecular Biology Laboratory, Meyerhofstraße 1, 69117 Heidelberg, Germany}
\author{Laeschkir Würthner\orcidlink{0000-0003-4302-4945}}
\affiliation{Developmental Biology Unit, European Molecular Biology Laboratory, Meyerhofstraße 1, 69117 Heidelberg, Germany}
\author{Isabella  R.\ Graf\orcidlink{0000-0001-9169-9109}}
\affiliation{Developmental Biology Unit, European Molecular Biology Laboratory, Meyerhofstraße 1, 69117 Heidelberg, Germany}
\affiliation{Department of Physics and Astronomy, Heidelberg University, 69120 Heidelberg, Germany}
%
\author{Caren Norden\orcidlink{0000-0001-8835-1451}}
\affiliation{Department of Physiology, Development and Neuroscience, Gurdon Institute, University of Cambridge, Tennis Court Road, CB2 1QN Cambridge, United Kingdom}
\author{L. Rondoni\orcidlink{0000-0002-4223-6279}}
\affiliation{Department of Mathematical Sciences, Politecnico di Torino, Corso Duca degli Abruzzi 24, 10129 Turin, Italy}
\affiliation{INFN, Sezione di Torino, Turin 10125, Italy}
\author{A. Erzberger\orcidlink{0000-0002-2200-4665}}%
\email{Corresponding author 1: erzberge@embl.de}
\affiliation{Cell Biology and Biophysics Unit, European Molecular Biology Laboratory, Meyerhofstraße 1, 69117 Heidelberg, Germany}
\affiliation{Department of Physics and Astronomy, Heidelberg University, 69120 Heidelberg, Germany}
\author{R. Belousov\orcidlink{0000-0002-8896-8109}}
\email{Corresponding author 2: roman.belousov@embl.de}
\affiliation{Cell Biology and Biophysics Unit, European Molecular Biology Laboratory, Meyerhofstraße 1, 69117 Heidelberg, Germany}



\date{\today}

\begin{abstract}
    Motivated by the complex material properties of living matter, we derive a framework of stochastic thermo- and elastodynamics for active materials from the classical principle of least action and the Rayleighian formalism for external forces. Our approach couples the local barycentric collective coordinate and the streaming velocity as the canonical conjugate pair, which unifies various descriptions of elastic and viscous soft matter, including Helfrich theory for lipid membranes, linear and nonlinear elasticity of thin shells and deformable droplets, as well as hydrodynamics of active and passive fluids. The Lagrangian mechanics of collective field variables permits direct applications of the results and theorems from statistical mechanics to existing models of complex soft and living matter. To illustrate the potential applications of stochastic elastodynamics in biology, we derive and simulate a boundary-driven deformation model of an elastic cell nucleus in the presence of fluctuations.
\end{abstract}

\maketitle

Living matter constitutes a class of materials with particularly complex properties, including viscoelasticity, anisotropy, and compositional heterogeneity~\cite{Fletcher2010,Marchetti2013,Chaudhuri2020,Takatori2020,Kirschbaum2021,Grauer2021,Peterson2021,Iyer2022,Rajendran2023,Fries2025,Ichikawa2026}. In addition, living systems are intrinsically out of equilibrium, with continuous energy dissipation coupled to feedback and adaptation \cite{Discher2005,Dullweber2025}.
Correspondingly, a wide range of theoretical frameworks---including continuum active hydrodynamics, vertex and Cellular Potts models, particle-based simulations, and phase-field approaches---have been developed to describe different aspects and scales of living systems.
Typically, these models are fit to specific research areas---e.g.\ related to surface and interface properties~\cite{Jlicher2018,Terzi2019,TorresSnchez2019,Hossein2020,Zhang2020,Shankar2021,TorresSnchez2022,Salbreux2022,Hecht2022,Foley2023,Link2023,Frey2024,Varma2025,Drozdowski2026}, dynamics and development of tissues, cells, and their organelles~\cite{TorresSnchez2021,Estabrook2021,Luciano2022,Thiagarajan2022,Bauermann2023,Takaki2023,LpezGuajardo2023,Bondarenko2023,Belousov2024,MaiaGil2024,Fuhrmann2024,Link2024,Lettermann2024,Vercruysse2024,McLaren2025},---and can be combined to describe more complex settings~\cite{Komrlj2017,Kelkar2022,Hirway2022,Guo2022,Dinet2023,Pnisch2023,Lu2024,PomboGarca2024,Brckner2024,Lange2025,Dessalles2025,Ouzeri2025,Bailleul2025,Keijzer2025,Holzapfel2025,Monfared2025,GomezMelo2026}. To systematically compare these approaches and synthesize physical theories across diverse systems, a unifying conceptual framework is needed.

Although thermodynamics and \textit{stochastic} thermodynamics~\cite{Peliti2021,Seifert2025} already provide such a framework for discrete-state, particle-level, or agent-based descriptions, their extension to hydrodynamics for fluctuating field variables is still an area of ongoing research~\cite{Belousov2018,Agranov2021,Markovich2021,Cates2022,Kodama2022,Rower2022,Wrthner2023,Yoshimura2024,Giona2025,Belousov2026}. To identify key principles that govern the physical processes of interest, while abstracting away from details that can be approximated by effective models, the reductionist perspective~\cite{Chibbaro2014}---immanent to these theories---appears particularly suitable.

The classical action principle provides a solid foundation for thermodynamics at the level of molecular motion. The exploitation of this foundation at the continuum level has however been limited. While Lagrangian mechanics and related formalisms successfully describe conservative elastodynamics~\cite{Simo1988,pb2026}, its extension for hydrodynamic equations of motion with dissipative phenomena proved to be surprisingly complicated~\cite{Eckart1960,Finlayson1972,Mobbs1982,Zakharov1997,deSaxc2024,Sanders2024,Botelho2025,Taha2023}. A thermodynamically consistent formalism for the stochastic dynamics of continuum active matter remains therefore yet incomplete.

In this paper we connect the thermodynamics of elastic and viscous continuum systems to the classical action principle. In particular, we extend the proof of the fluctuation-dissipation theorem for such systems within the linear-response limit.
Building on the paradigmatic example of the harmonic oscillator and its successful variational treatment in stochastic thermodynamics, we first develop the mechanical action principle for \textit{stochastic elastodynamics}, taking advantage of the analogy between the theories of Hookean springs and linear elastic materials. Moreover, following this approach in a more general setting, we construct a comprehensive framework of stochastic thermodynamics for more complex elastic interactions, as well as for hydrodynamic systems.

In the variational formulation of classical mechanics we consider an $N_d$-dimensional system described by time-dependent variables $\big(\bm{q}(t), \dot{\bm{q}}(t)\big)$, which for coupled harmonic oscillators include all the coordinates $\bm{q}_{i=1,2,...}$ and velocities $\dot{\bm{q}}_i$, whereas for an elastic body they refer to the components of a displacement vector $\bm{q}(t) = \bm{u}(t,x_{i=1,...,N_d})$ and its time derivative $\dot{\bm{q}}(t) = \partial_t \bm{u}$. Throughout this work, lowercase $x_i$ denotes continuous spatial coordinates of a field theory, while uppercase $X_i$ refers to discrete particle positions. We represent the dynamics of the system state as the Euler-Lagrange-Rayleigh equations, which in the vector form read
\begin{equation}\label{eq:archetype}
    \frac{\delta\action[\bm{q}(t)]}{\delta\bm{q}} = \frac{\delta\rayleigh}{\delta\dot{\bm{q}}},
\end{equation}
where $\action$ is the \textit{action} functional varied with respect to path $\bm{q}(t)$ and $\rayleigh$ is the Rayleigh \textit{dissipation} potential differentiated or varied with respect to velocity $\dot{\bm{q}}$. \textit{Local optimization} of a generalized dissipation potential with respect to $\dot{\bm{q}}$ also inspired a related Onsager principle~\cite{Doi2011,Arroyo2017,Wang2021,Cho2021,Beris2024}, which should be distinguished from \textit{global optimization} of the classical mechanical action with respect to the \text{path} $\bm{q}(t)$.

Equation~\eqref{eq:archetype} provides a unified basis for deriving both elastic and hydrodynamic theories by combining a Lagrangian action for conservative stresses with a Rayleighian description of dissipation, fluctuations, active stresses, and external driving. The potential energy of such systems depends on their configuration described by the displacement field $\bm{u}$ either explicitly, e.g. in Hookean elasticity, or—as we show here for hydrodynamic systems—\textit{implicitly}. In both cases the metric tensor, which encodes deformations of the body reference geometry, plays the central role. The Hookean model of linear elasticity is derived through expansion of the metric tensor in $\bm{\nabla u}$ to the first order. The hydrodynamic equations of motion incorporate the geometric effects of local deformations through the fluid density. Moreover, our analysis of the microscopic connection between the macroscopic fluid dynamics and molecular motion identifies the displacement field $\bm{u}$ with a local center of mass. As a collective coordinate, it is the canonical conjugate of the local momentum density, wheres its time derivative represents the streaming velocity $\bm{v} = \partial_t \bm{u}$.

Once the dynamics of an elasto- or hydrodynamic system is expressed in the form of \cref{eq:archetype}, the analysis of energetics and thermodynamics becomes a routine task. By identifying the Hamiltonian, incorporating active forces, fluctuations, and dissipation, we extend the standard formalism of stochastic thermodynamics to such systems, and show that in the limit where the size of an elastic body is negligible, we recover the Langevin equation for a Brownian particle.

The Hookean potential energy can be readily replaced with other expressions used in models of soft matter, as discussed in~\cref{sec:disc}. Our framework then provides a systematic way to derive equations of motion, to incorporate nonconservative forces, e.g. active forces and fluctuations, and to obtain a consistent thermodynamic interpretation.

As a numerical example, we derive and simulate the stochastic elastodynamics of a cell nucleus, which has been shown to act as a mechanosensor of the surrounding cellular environment in diverse biological contexts \cite{Kirby2018,Lomakin2020,Kalukula2022}. Specifically, we consider a passive elastic inclusion in an active environment, subject to pulling and pushing forces, spontaneous fluctuations, and friction with the medium. In the overdamped limit---as appropriate at this scale---we obtain dynamics, which are entirely determined by the system's boundary degrees of freedom and, therefore, can be reduced to a numerical evolution of the nucleus surface. We quantify how the resulting shape features and volume dynamics encode information about force profiles at the biophysical scales of cytoskeletal processes, and present possible generalizations.

\section{\label{sec:action}Action principle}
First we revisit the action principle for isolated elastic systems, which we extend to include external forces of various kinds in \crefrange{sec:act}{sec:ext}. This section concludes with a general stochastic thermodynamic theory of elastic materials, followed by hydrodynamic theory in \cref{sec:hydro} and numerical examples in \cref{sec:num}. Discussion of the action principle for other systems is provided in \cref{sec:disc}.

To evaluate the standard form of the action functional
\begin{equation}\label{eq:action}
    \action[\bm{q}(t)] = \int_0^{\mathcal{T}} dt\, \mathcal{L}\big(\bm{q}(t), \dot{\bm{q}}(t)\big),
\end{equation}
we must specify the \textit{Lagrangian} $\mathcal{L}(\bm{q}, \dot{\bm{q}})$ of the system, which consists of the kinetic and potential terms. For example, the Lagrangian of a one-dimensional harmonic spring of stiffness $K$ and rest length $L$, which connects two particles $\bm{q} = (X_1, X_2)$ of masses $M_{i=1,2}$, reads
\begin{equation}\label{eq:lag_osc}
    \mathcal{L}_{1:2} = \sum_{i=1,2} \frac{M_i  \dot{X}_i^2}{2} - \frac{K}{2} (X_2 - X_1 - L)^2.
\end{equation}
Here, the subscript $1:2$ indicates that we are working with a system or subsystem composed of particles $1$ and $2$.
In the standard manner (\cref{app:osc}), the principle of least action yields the Newtonian equations of motion for the two particles, the canonical momenta $\bm{p} = (p_1, p_2)$ and the conserved total energy $\mathcal{H}_{1:2}$.

\begin{figure}[!t]\centering
\includegraphics[width=\columnwidth]{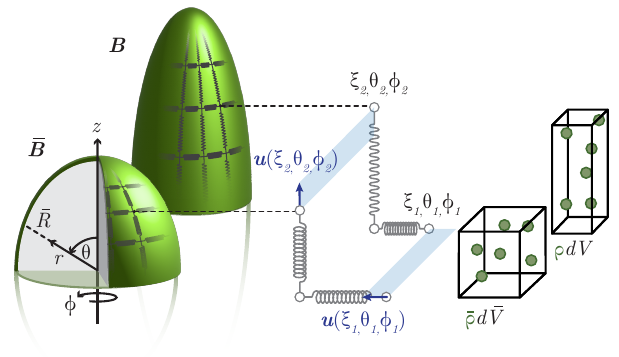}
\caption{\label{fig:ref}Displacement vector of a body $\bm{B}$ with respect to a reference configuration $\bar{\bm{B}}$, which are isomorphic to a unit ball $(\xi,\theta,\phi)$ (\cref{app:ref}). For illustration, when the boundary surface $\partial \bar{\bm{B}}$ can be described in spherical coordinates $(r,\theta,\phi)$ as $\partial B^r = \bar{R}(\theta,\phi)$ associated to the geometric center of the body, this isomorphism may be represented in a particularly simple form $(B^r, B^\theta, B^\phi) = \big(\bar{R}(\theta,\phi)\xi, \theta,\phi\big)$. Once the points of $\bm{B}$ and $\bm{\bar{B}}$ are charted using $(\xi,\theta,\phi)$, they can be transformed to any convenient reference frame, e.g. Cartesian. Then the displacement $\bm{u}(\xi,\theta,\phi)$ maps each point of the reference configuration $\bar{\bm{B}}$ to the deformed body, ${\bm{B}}$, $\bm{B} = \bar{\bm{B}}(\xi,\theta,\phi) + \bm{u}(\xi,\theta,\phi)$. Each point of an ideal elastic solid carries a conserved mass $d\bar{V}\bar\rho = dV \rho$~\cite[Sec.~5.5.2]{irgens2008continuum} and can be thought of as being connected by tiny springs to its nearest neighbors. The displacement field, except for one generating a rigid-body motion, deforms neighborhoods of these points, thus extending and compressing the springs.
}
\end{figure}

Whereas the continuum generalization of \cref{eq:lag_osc} to an infinite elastic medium is straightforward (\cref{app:lim}, Ref.~\cite[Sec.~1.2]{kardar2007statistical}), the case of a finite body bounded in three-dimensional space---relevant to many applications---necessitates an adaptation of the classical theory. Here we focus in particular on bodies enclosed by a surface \textit{topologically equivalent} to a sphere (these include cells, their nuclei and some other organelles). Each point of such a body $\bm{B}(\xi, \theta, \phi)$ can be mapped onto a unit ball parameterized by three dimensionless coordinates: a rescaled radius $\xi \in [0,1]$, the polar angle $\theta \in [0, \pi]$, and the azimuthal angle $\phi \in [0, 2\pi)$. We choose the displacement vector $\bm{u}(\xi,\theta,\phi) = \bm{B} - \bar{\bm{B}}$ as the generalized coordinate $\bm{q}$, with respect to some \textit{reference configuration} of the system $\bar{\bm{B}}$ (\cref{fig:ref,app:ref,app:lim}). In the following, we denote quantities defined in the reference configuration with an overbar.

The triplet of \textit{internal} or \textit{material} coordinates $(\xi, \theta,\phi)$ plays the same role as the index $i$ in \cref{eq:lag_osc}, which simply identifies a material point by exploiting the isomorphism $\bm{B}(\xi,\theta,\phi)$. The summation operator can be replaced in the continuum limit by an integral over the volume $\bar{V}$ of the whole body in the reference configuration:
\begin{multline}\label{eq:lag_ela}
    \mathcal{L}_\ela = \int_{\bar{V}} d\bar{V} \left[\frac{\bar\rho (\partial_t\bm{u})^2}{2} - \mathcal{U}_\ela\right] \\= \int_{\bar{V}} d\bar{V} \frac{\bar\rho (\partial_t\bm{u})^2}{2}
    - U_\ela[\bm{\nabla u}],
\end{multline}
where $\bar\rho$ is the mass density of the reference configuration, while $U_\ela$ and $\mathcal{U}_\ela$ are respectively the elastic energy and its density.

As discussed in the following, potentials of the form $U[\bm{\nabla u}]$, which depend on the gradient tensor $\bm{\nabla u}$, are of particular interest, both in elastodynamics and, indirectly through the fluid density, in hydrodynamics. Within the domain of Hookean elasticity regarding quadratic potentials in $\bm{\nabla u}$, we focus here on an isotropic homogeneous body, whose energy density can be described in three dimensions by just two \textit{constant} Lam\'e coefficients $\bar\lambda$ and $\bar\mu$~\cite[Chapter I]{landau2005theory}, \cite[Sec.~7.2]{irgens2008continuum}:
\begin{equation}\label{eq:hooke_ela}
    \mathcal{U}_\ela(\bm{\nabla u}) = \frac{\bar\lambda}{2} (\tr\bm{\epsilon})^2 + \bar\mu \fr{\bm{\epsilon}}^2,
\end{equation}
where the operators $\tr$ and $\fr{\cdot}$ stand for the trace and the coordinate-invariant Frobenius norm of the \textit{linear strain} tensor $\bm{\epsilon} = (\bm{\nabla u} + \tra{\bm{\nabla u}}) / 2$ (\cref{app:strain,app:hooke}). The coefficients $\bar\mu$ and $\bar\lambda$ are intrinsic material properties: $\bar\mu$ is the shear modulus, resisting shape-changing deformation, while their combination $\bar\lambda + 2\bar\mu/3$ defines the bulk modulus, resisting volumetric compression. Note that \cref{eq:hooke_ela} implies that the reference configuration $\bar{\bm{B}}$ corresponds to the minimum of the potential $U_\ela[\bm{\nabla u}]|_{\bm{u}\equiv\bm{0}}$.

By extremizing the action functional \cref{eq:action} with the Lagrangian given by \cref{eq:lag_ela,eq:hooke_ela}, we get the Euler-Lagrange equations (\cref{app:stress,app:eul_lag_hooke})
\begin{equation}\label{eq:daction_ela}
    \bm{0} = \frac{\delta \action_\ela}{\delta\bm{u}} = -\frac{d}{dt}(\sqrt{\bar{g}}\bar{\rho} \partial_t\bm{u}) + \sqrt{\bar{g}} \bm{\nabla} \cdot \bm{\sigma}_\ela,
\end{equation}
where the metric tensor $\bar{g}=\det \bar{\bm{g}}$ encodes how lengths scale in the chosen coordinate system, and $\bm{\sigma}_\ela = \partial\mathcal{U}_{\ela} / \partial \bm{\nabla u}$ is the elastic stress, representing the thermodynamic force conjugate to the displacement gradient in this formalism. Equation~\eqref{eq:daction_ela} implies a traction-free boundary condition \cref{eq:zero_traction}.

Because the mass element $\sqrt{\bar{g}} \bar\rho$ of a perfect elastic solid is conserved (\cref{app:mass}, \cref{fig:ref}, Ref.~\cite[Sec.~5.5.2]{irgens2008continuum}), \cref{eq:daction_ela} is equivalent to the Cauchy equation of motion\footnote{
    Body forces can be included in the potential by additional terms which we do not discuss for simplicity.
}~\cite[Sec.~3.2.5]{irgens2008continuum}
\begin{equation}\label{eq:cauchy}
    \bar\rho\partial_t^2 \bm{u} = \bm{\nabla}\cdot\bm{\sigma}_\ela.
\end{equation}

Given the Lagrangian~\eqref{eq:lag_ela}, we can define the momentum density
\begin{equation}\label{eq:mom_ela}
    \bm\pi = \frac{\delta \mathcal{L}_\ela}{\delta\partial_t\bm{u}} = \bar\rho\partial_t\bm{u}
\end{equation}
conjugate to the displacement coordinate $\bm{u}$, and the Hamiltonian density
\begin{equation}\label{eq:ham_ela}
    \mathcal{H}_\ela = \int_{\bar{V}} d\bar{V} \bm\pi \cdot \partial_t\bm{u} - \mathcal{L}_\ela = \int_{\bar{V}} d\bar{V} \frac{\bar\rho (\partial_t \bm{u})^2}{2} + U_\ela.
\end{equation}
Hence, the Lagrangian~\eqref{eq:lag_ela}, the Cauchy equation of motion \eqref{eq:cauchy}, the momentum density \eqref{eq:mom_ela}, and the Hamiltonian \eqref{eq:ham_ela} of an elastic body follow the same structure as the Lagrangian \eqref{eq:lag_osc}, the Newtonian equations of motion~\eqref{eq:newt_osc}, the momentum \eqref{eq:mom_osc}, and the Hamiltonian~\eqref{eq:ham_osc} of the harmonic oscillator, respectively.

For reference configurations and parameters, which are constant in time, such as $\bar\lambda$ and $\bar\mu$, the Hamiltonian given by Eq.~\eqref{eq:ham_ela} is conserved by the equation of motion \cref{eq:cauchy}. In particular, from \cref{eq:ham_ela} it follows that
\begin{multline}\label{eq:conservation_ela}
    \frac{d\mathcal{H}_\ela}{dt} = \int_{\bar{V}} d\bar{V}  \left(\bar\rho \partial_t \bm{u} \cdot \partial_t^2\bm{u} + \bm{\sigma}_\ela \cdot \bm{\nabla}\partial_t\bm{u}
    \right)
    \\= \int_{\bar{V}} d\bar{V}\,  \partial_t \bm{u} \cdot \left(\bar\rho \partial_t^2\bm{u} - \bm{\nabla} \cdot \bm{\sigma}_\ela
    \right) \equiv 0,    
\end{multline}
in which we used integration by parts and the Cauchy \cref{eq:cauchy} with the natural boundary condition \eqref{eq:zero_traction}.

In the following, we incorporate active and stochastic stresses into \cref{eq:conservation_ela}. Importantly, our framework permits describing both thermal and active stochastic processes. Thermal fluctuations constitute a special case, in which the stochastic forcing satisfies the fluctuation-dissipation theorem and the stationary distribution reduces to the Boltzmann distribution
\begin{equation}\label{eq:boltzmann}
    p(\bm{u}, \dot{\bm{u}})\propto e^{- \beta \mathcal{H}(\bm{u}, \dot{\bm{u}})},
\end{equation}
where $\beta = (\kB T)^{-1}$ with the Boltzmann constant $\kB$ and temperature $T$.

In a more general setting, which we consider below, the Hamiltonian may also contain an explicit time dependence.

\subsection{\label{sec:act}Active systems}
The action principle also applies to nonconservative systems with Hamiltonians explicitly depending on time and, therefore, not conserved. This time dependence can be introduced through the system's parameters, such as the Lam\'e coefficient or the rest extension, for example to model active processes. Indeed, living matter often exhibits active changes of its preferred configuration or its elastomechanical properties~\cite{Andreu2021,Skory2023,Alraies2024}.

Returning to the example of a simple harmonic oscillator described by the Lagrangian \eqref{eq:lag_osc}, one could introduce a time-dependent stiffness $K \to K_t$ or a rest extension $L\to L_t$---both these modifications can be encountered in applications of stochastic thermodynamics~\cite{Schmiedl2007,Blaber2022,Loos2024}. For convenience and without loss of generality, we consider the rest extension of the spring in the form $L_t = \bar{L} + F_\act(t) / K_t$ with a constant extension $\bar{L}$ and an active force $F_\act$. This parametrization allows us to cast the potential energy \cref{eq:pot_osc} as
\begin{equation}\label{eq:pot_t_osc}
    U_t = \frac{K_t}{2}(X_2 - X_1 - \bar{L})^2 - F_\act (X_2 - X_1 - \bar{L}) + \cancel{\text{o.t.}},
\end{equation}
where the subscript $t$ indicates the explicit time dependence, and the offset term $\text{o.t.} = F_\act^2 / (2 K_t)$ is ignored, because it has no effect on the equations of motion~\eqref{eq:newt_osc}. The complete differential of the Hamiltonian \cref{eq:ham_osc} then becomes
$$
    \frac{d\mathcal{H}_t}{dt} = \frac{\partial U_t}{\partial K_t}  \frac{dK_t}{dt} + \frac{\partial U_t}{\partial F_\act} \frac{dF_\act}{dt} = \frac{\dbar W_K}{dt} + \frac{\dbar W_F}{dt},
$$
in which $\dbar{W}_K$ and $\dbar{W}_F$ denote the work done on the system by active processes $K_t$ and $F_\act$, whereas the terms arising from $\partial H_t / \partial X$ and $\partial H_t / \partial\dot{X}$ are canceled out as per \cref{eq:conservation_osc}. The struck $d$, commonly used in classical thermodynamics, denotes here the path-dependent quantities, which by themselves represent inexact differentials. In stochastic thermodynamics one usually refers to energy flows represented by power $\dot{W}_{K,L} = \dbar{W}_{K,L}/dt$.

To model active mechanisms within the framework of Hookean elasticity, we transform the potential \cref{eq:hooke_ela} into (\cref{app:act})
\begin{equation}\label{eq:pot_act}
    \mathcal{U}_\act(\bm{\nabla u}) = \frac{\bar\lambda_t}{2} (\tr\bm{\epsilon})^2 + \bar\mu_t \fr{\bm{\epsilon}}^2 + \bm{\sigma}_\act:\bm{\nabla u},
\end{equation}
in which the time-dependent Lam\'e coefficients $\bar\lambda_t$ and $\bar\mu_t$ are analogues of the time-dependent stiffness $K_t$ [cf. \cref{eq:lim_act}], whereas the \textit{active stress} $\bm{\sigma}_\act$ generalizes the mechanism described by the active force $F_\act$ in \cref{eq:pot_t_osc}, with the colon denoting the double-index tensor contraction. Indeed, the principle of least action with the potential \cref{eq:pot_act} gives (\cref{app:eul_lag_hooke})
\begin{equation}\label{eq:cauchy_act}
    \bar\rho\partial_t^2 \bm{u} = \bm{\nabla}\cdot\big(
        \bm{\sigma}_\ela(t) + \bm{\sigma}_\act
    \big),
\end{equation}
where the elastic stress $\bm{\sigma}_\ela(t)$ may depend explicitly on time through the Lam\'e coefficients.

The temporal and spatial dependence of Lam\'e coefficients and the active stress in \cref{eq:cauchy_act} can be specified as various driving \textit{protocols}. For instance, these properties can be explicitly given as a priori known deterministic functions or defined implicitly by equations including stochastic or feedback terms that depend on the system's state~\cite{Schmiedl2007,Blaber2022,Loos2024,Thipmaungprom2026}.

Importantly, the action principle imposes one restriction on the time-dependent parameters of the system: they must not be \textit{explicit} functions of the state variables (\cref{sec:ext}). Such external driving forces are instead incorporated into the extension of the action-principle framework discussed below.

\subsection{\label{sec:flu}Dissipation and fluctuations}
The Rayleighian dissipation potential provides a widely adopted framework that extends the variational formulations of classical mechanics to include external forces~\cite{Lookman2003,Minguzzi2015}, \cite[Sec.~7.1]{Doi2013}. With a slight generalization it can also incorporate nonconservative state-dependent forces, enabling e.g.\ the modeling of spontaneous fluctuations. The prototype of this formalism is the Rayleigh dissipation function~\cite[Sec.~1.5]{Goldstein2012} with a friction coefficient $\gamma_0$:
\begin{equation}\label{eq:ray_proto}
    \rayleigh_0(\dot{\bm{q}}) = \frac{\gamma_0 \dot{\bm{q}}^2}{2},
\end{equation}
whose derivative with respect to the velocity $\dot{\bm{q}}$ is typically equated to the variation of action. In the case of a harmonic oscillator, for instance, by using \cref{eq:dS_osc_first,eq:dS_osc_second} with
\begin{equation}\label{eq:eul_lag_rel_osc}
    \frac{\delta \action_{1:2}}{\delta X_i} = \frac{\partial\rayleigh_{1:2}}{\partial\dot{X}_i},
\end{equation}
we can introduce dissipative forces that contain particle-specific friction coefficients $\gamma_i$ for each degree of freedom in the Rayleigh potential $\rayleigh_{1:2} = \sum_i \gamma_i \dot{X}_i^2 / 2$. The latter follows the same structure as its prototype $\rayleigh_0$ and entails linear nonconservative terms in the Newtonian equations of motion~\eqref{eq:newt_osc}:
$$
    M \ddot{X}_i = - \frac{\partial U_{1:2}}{\partial X_i} - \gamma_i \dot{X}_i.
$$

The more recent foundations of the Rayleigh formalism~\cite{Lookman2003,Minguzzi2015}, \cite[Sec.~7.1]{Doi2013} encompass a more comprehensive interpretation of \cref{eq:archetype,eq:eul_lag_rel_osc}, the two sides of which refer to parts of a complete action for a supersystem comprising the system of interest and an external bath. Indeed, let us split the Lagrangian~\cref{eq:lag_osci} of the harmonic chain into two parts $\mathcal{L}_{1:N} = \mathcal{L}_{1:2} + \Delta\mathcal{L}_{2:N+1}$ where $\mathcal{L}_{1:2}$ coincides with~\cref{eq:lag_osc} for the harmonic oscillator of stiffness $K = K_1$, while $\Delta\mathcal{L}_{2:N+1}$ contains the remaining terms. By minimizing the action \cref{eq:action} with respect to $(X_1, X_2)$, we then obtain two contributions associated with these parts of the full Lagrangian adding up to zero:
\begin{equation}\label{eq:split_osc}
    \frac{\delta \action_{1:2}}{\delta X_i} + \frac{\delta \Delta\action_{2:N+1}}{\delta X_i} = 0,
\end{equation}
which has the structure of \cref{eq:eul_lag_rel_osc} if we can find a dissipation potential $\rayleigh_{1:2}$, such that
\begin{equation}\label{eq:ray}
    \frac{\partial\rayleigh_{1:2}}{\partial\dot{X}_i} \simeq -\frac{\delta\Delta\action_{2:N+1}}{\delta X_i}.
\end{equation}

The problem stated by \cref{eq:ray} is most generally addressed within the Mori-Zwanzig theory~\cite{teVrugt2019,teVrugt2020}, \cite[Sec.~15.7]{Tuckerman2023}, which beyond the dissipative terms leads to the fluctuation-dissipation theorem and friction kernels with memory effects. Although in practice this formalism almost always requires an approximate treatment, the harmonic interactions allow for an exact formal solution due to the linear character of the resulting Euler-Lagrange equations~\cite{Deutch1971}, \cite[Chapter 15]{Tuckerman2023} (\cref{app:fdt_chain}).

Generalized for the case of Hookean elasticity (\cref{app:fdt_hooke}), a simple form of the Rayleigh potential in the white-noise limit reads
\begin{equation}\label{eq:ray_delta_wave}
    \rayleigh_\ela = \int_{\partial\bar{V}} dS\ \left(\frac{\bar\gamma\partial_t\bm{u}^2}{2} - \sqrt{2\kB T \bar{\gamma}}\ \partial_t\bm{u} \cdot \partial\bm{w}\right),
\end{equation}
with a surface-bound friction constant $\bar\gamma$, a spatially-distributed standard white noise $\partial\bm{w}(\theta, \phi)$~\cite{Belousov2018} defined on the boundary $\partial\bar{V}$, and the Stratonovich interpretation of the dot product. By varying $\rayleigh_\ela$ with respect to $\partial_t\bm{u}$, the surface integral can be converted into a volume integral as in \cref{eq:space3}, to obtain the Langevin-Cauchy equations of motion
\begin{equation}\label{eq:langevin_cauchy}
    \bar\rho\partial_t^2 \bm{u} + \bar{\gamma} \frac{\bar{n}}{\sqrt{\bar{g}}}\delta(\xi-1) \partial_t\bm{u} - \bm{\nabla}\cdot(\bm{\sigma}_\ela + \bm{\sigma}_\act) = \bm{f}_\flu.
\end{equation}
where the delta-correlated---in space and time---Gaussian random traction force density $\bm{f}_\flu$
is
$$ \bm{f}_\flu = \sqrt{2\kB T \bar{\gamma}}\ \frac{\bar{n}} {\sqrt{\bar{g}}}\delta(\xi-1)\partial\bm{w}, $$
and $\bar{n}$ is the magnitude of the signed-area vector~\cref{eq:normal}.

The fluctuation and dissipation forces do work on the system, which can be considered as heat $\dot{Q}$ exchanged with the bath. The first law of thermodynamics can be directly inspected by the time derivative of the Hamiltonian:
\begin{align}
    \label{eq:thermo_act}
    &\dot{H}_\act = \dot{Q} + \dot{W}_\act,
    \\\label{eq:Qdef}
    &\dot{Q} = -\int_{\partial\bar{V}} dS\, \left(\bar\gamma\partial_t\bm{u}^2 - \sqrt{2\kB T \bar{\gamma}}\ \partial_t\bm{u} \cdot \partial\bm{w}
    \right),\\
    &\dot{W}_\act = \int_{\bar{V}} d\bar{V}\,  \partial_t \bm{u} \cdot \bm{\nabla} \cdot \bm{\sigma}_\act.
\end{align}
More complex versions of the fluctuation-dissipation terms, which include the hydrodynamic viscous friction---not bound to the surface---and memory kernels are also discussed in \cref{app:fdt_hooke}.

\subsection{\label{sec:ext}Externally driven systems}
To complete our framework, we consider external nonconservative forces, which can be incorporated directly into the Rayleighian and maintain the system out of equilibrium. While physically distinct from material-intrinsic processes described through the active-stress term in \cref{sec:act}, some forms of such an external driving may lead to an equivalent equation of motion. Nonetheless, there is a particular scenario where the nonconservative forces unambiguously fall into the dominion of the Rayleighian potential: feedback forces that explicitly depend on the state variables. It is most easy to illustrate by adding an external driving $\bm{F}^\ext = (\Delta K X_2, 0)$ to equations of motion \eqref{eq:newt_osc}, which become
\begin{align}\label{eq:osc_ext1}
    M \ddot{X}_1 = K X_2 - K X_1 - K L + \Delta K X_2,
    \\\label{eq:osc_ext2}
    M \ddot{X}_2 = -K X_2 + K X_1 + K L.
\end{align}
In the above equations, the energy~\cref{eq:ham_osc} is not conserved by \cref{eq:osc_ext1,eq:osc_ext2}:
$$
    \frac{dH_{1:2}}{dt} = \dot{X}_1 F^\ext_1 = \Delta K \dot{X}_1 X_2 = \dot{W}_\ext, 
$$
where $\dot{W}_\ext$ is the external power supplied to the system.

Importantly, there is no way to incorporate this force into the Lagrangian, because the right-hand sides of \cref{eq:osc_ext1,eq:osc_ext2} do not form an exact differential. For instance, if we modify the potential as $U + \Delta K X_1 X_2$, leading to the term $F^\ext_1$ in \cref{eq:osc_ext1}, then it must also generate a missing back-action term $- \Delta K X_1$ in \cref{eq:osc_ext2}.

The Rayleigh potential provides a straightforward way to incorporate external driving, such as the above force $F^\ext$. Indeed we can add another term
\begin{equation}\label{eq:ray_ext_osc}
    \Delta\rayleigh^\ext_{1:2} = -\sum_i F^\ext_i(t, X_1, X_2) \dot{X}_i
\end{equation}
to the Rayleighian $\rayleigh_{1:2}$ that would generate from \cref{eq:ray} the force terms $F_{\ext, 1}$ and $F_{\ext,2}\equiv0$ on the right-hand sides of \cref{eq:osc_ext1,eq:osc_ext2}.

Generalization of \cref{eq:ray_ext_osc} for elastodynamic systems is trivial:
\begin{equation}
    \Delta\rayleigh_\ext = -\int_{\bar{V}} d\bar{V}\ \partial_t \bm{u} \cdot \bm{\nabla}\cdot\bm{\sigma}_\ext,
\end{equation}
leading to the driven Langevin-Cauchy equations of motion
\begin{equation}\label{eq:most_general_wave}
    \bar\rho\partial_t^2 \bm{u} + \bar{\gamma}\frac{\bar{n}}{\sqrt{\bar{g}}}\delta(\xi-1) \partial_t\bm{u} - \bm{\nabla}\cdot(\bm{\sigma}_\ela + \bm{\sigma}_\act + \bm{\sigma}_\ext) = \bm{f}_\flu,
\end{equation}
with the first law of thermodynamics~\eqref{eq:thermo_act} extended by
\begin{align}\label{eq:first_law_ext}
    \dot{H}_\act =& \dot{Q} + \dot{W}_\act + \dot{W}_\ext,\\
    \dot{W}_\ext =& \int_{\bar{V}} d\bar{V}\ \partial_t \bm{u} \cdot \bm{\nabla}\cdot\bm{\sigma}_\ext.
\end{align}

When the external forces are not of the feedback type, as discussed above, they can be included as external potentials.
Although the effects of such modifications are quite straightforward, even when they are not restricted to harmonic interactions, for brevity we are not discussing such contributions further.

\subsection{\label{sec:thermo}Thermodynamics in the overdamped limit}
In many living systems, dissipation dominates over inertia on the relevant time and length scales, rendering the acceleration term in the equations of motion negligible. 
Therefore, while \cref{eq:first_law_ext} offers a complete foundation for the stochastic thermodynamics of elastic materials, we proceed by developing its overdamped limit.

Again,
in complete analogy to the overdamped model of a harmonic oscillator, we eliminate the acceleration from \cref{eq:most_general_wave} in the limit $\bar\rho / \bar{\gamma} \to 0$ and obtain
\begin{equation}\label{eq:overdamped}
\frac{\bar{n}}{\sqrt{\bar{g}}}\delta(\xi-1) \bar{\gamma} \partial_t\bm{u} = - \frac{\delta U_\act}{\delta \bm{u}} + \bm{\nabla}\cdot\bm{\sigma}_\ext + \bm{f}_\flu.
\end{equation}
Together with this transformation we should eliminate the kinetic term in the Hamiltonian~\eqref{eq:ham_ela}. Then the first law of stochastic thermodynamics in the overdamped limit reads
$$
    \dot{U} = \dot{Q} + \dot{W}_\ext + \dot{W}_\act,
$$
where $U$ refers to the potential energy of the system.

Continuum theory describes field dynamics that vary on sufficiently large scales to permit the coarse-graining of microscopic dynamics---a separation of scales that necessarily arises for densities of conserved quantities and other hydrodynamic variables~\cite[Sec. 8.6]{ChaikinLubensky1995Hydrodynamics}. At such macroscopic scales, the effects of thermal fluctuations typically become negligible~\cite{Belousov2018}. While very small particles ceaselessly exhibit Brownian motion---in fact our theory is fully consistent with this limit of small bodies (\cref{app:particle}),---under the effect of friction, larger bodies come to a full arrest or move with a constant terminal velocity. However, in active materials including living matter, stochastic nonequilibrium processes can produce fluctuations at the macroscopic scale, which relevantly interact with the slow hydrodynamic variables. The random force density $\bm{f}_\flu$ in [add eq. ref] permits describing the stochastic dynamics of such systems.


\section{\label{sec:hydro}Hydrodynamic action}
There is a recurring interest in extending the classical action principle to the Euler and Navier–Stokes equations for fluid dynamics~\cite{Eckart1960,Finlayson1972,Mobbs1982,Zakharov1997,Taha2023,deSaxc2024,Banerjee2024,Sanders2024,Botelho2025}. The dominant idea originates from the theory of potential flows, in which the fluid velocity field $\bm{v} = \bm{\nabla} \varphi$ is related to the gradient of a scalar potential $\varphi$. However, because the velocity potential provides a single coordinate-like field $\varphi$, whereas the system has three independent components of $\bm{v}$, the relations between these quantities obfuscate the analogy with the classical pairs of conjugate coordinates and momenta. Moreover, that the velocity potential is not itself a directly observable state variable limits the direct applicability of the ensuing formalism. Second, when the velocity is interpreted as gradient of a scalar function, it cannot describe flows with nonzero vorticity $\bm{\nabla}\times\bm{v} \ne \bm{0}$, because the curl of a gradient vanishes identically $\bm{\nabla}\times\bm{\nabla}\phi \equiv \bm{0}$. This limitation can be partly remedied by adopting the so-called Clebsch transformation~\cite{Mobbs1982} to represent more complex configurations of $\bm{v}$ with help of two additional scalar parameters. However, it has not been shown to provide a general form of a vector field like, for example, the Helmholtz decomposition. Therefore the question remains open whether the Clebsch transformation is sufficient to capture all possible solutions of the hydrodynamic equations of motion. Moreover, the two additional scalar parameters render the theory even more abstract.

Our variational approach encompasses a more immediate extension of the classical action principle to fluid mechanics, where $\bm{u}$ is the coordinate conjugate to the momentum $\bm\pi = \rho\bm{v}$ with $\bm{v} = \partial_t\bm{u}$. Moreover, we do not seek to incorporate the dissipative contributions into the action, as some of the aforementioned authors suggest~\cite{Finlayson1972,Mobbs1982,Zakharov1997,deSaxc2024,Sanders2024,Botelho2025}. In the thermodynamic interpretation of our theory, the two sides of \cref{eq:archetype} are attributed to processes \textit{internal} and \textit{external} to the system, respectively.

\subsection{Variational formulation of hydrodynamics}
The variational principle outlined in \cref{sec:action} applies immediately to fluids if we replace the elastic potential $U_\ela$ with a general functional
$$
    U_\rho[\bm{\nabla u}] = \int_{\bar{V}} d\bar{V}\, \mathcal{U}_\rho(\rho) = \int_{\bar{V}} d\bar{V}\, \mathcal{U}_\rho\left(\frac{\sqrt{\bar{g}}\bar\rho}{\sqrt{g(\bm{\nabla u})}}\right),
$$
which depends directly on the density $\rho$ of the instantaneous configuration $\bm{B}$ (\cref{fig:ref}), and by virtue of \cref{eq:rho} in \cref{app:mass} implicitly on $\bm{\nabla u}$.
As in \cref{eq:daction_ela,eq:cauchy}, here we also assume $\bar\rho$ time-independent by neglecting diffusive and other transport phenomena. Note that to vary the action we continue to use the Eulerian picture, where the material coordinates refer to fixed points of the reference configuration moving in space with the streaming velocity (\cref{app:ref}). The variation of the potential energy then yields
\begin{multline}\label{eq:pressure}
    \delta U_\rho = \int_{\bar{V}} d\bar{V} \frac{\partial\mathcal{U}_\rho}{\partial\rho} \sqrt{\bar{g}}\bar\rho \delta g^{-\frac{1}{2}}
    = -\int_{\bar{V}} d\bar{V} \frac{\partial\mathcal{U}_\rho}{\partial\rho} \frac{\sqrt{\bar{g}}}{\sqrt{g}}\bar\rho\delta\bm{\nabla}\cdot{\bm{u}}
    \\= -\int_{\bar{V}} d\bar{V} p \bm{\nabla}\cdot\delta\bm{u}
    = \int_{\bar{V}}\ d\bar{V} \bm{\nabla} p \cdot \delta\bm{u}
\end{multline}
in which $p = \rho \partial{\mathcal{U}_\rho}/\partial\rho$ is the pressure.

By using \cref{eq:pressure} for $\delta U_\ela$ in the variation of action, we obtain the Euler equations of motion for fluids in the Eulerian reference frame
\begin{equation}\label{eq:eul_eul}
    \bar{\rho} \partial_t \bm{v}_\Eul + \bm{\nabla} p = 0,
\end{equation}
or in the Lagrangian picture (\cref{app:ref})
\begin{equation}\label{eq:eul_lag}
    \rho \left( \partial_t \bm{v}_\Lag + \bm{v}_\Lag\cdot\bm{\nabla v}_\Lag \right) + \bm{\nabla}p = 0.
\end{equation}
Finally, by using the Rayleighian~\cite[Sec.~7.1]{Doi2013}[cf. \crefrange{eq:ray_memo_wave}{eq:hydro} in \cref{app:fdt_hooke}]
\begin{equation}\label{eq:ray_hydro}
    \tilde{\rayleigh} = \int_{\bar{V}} d\bar{V} \left(\frac{\tilde\lambda}{2} (\tr\dot{\bm{\epsilon}})^2 + \tilde\mu \fr{\dot{\bm{\epsilon}}}^2
    \right)
\end{equation}
with viscous Lam\'e coefficients $\tilde\lambda$ and $\tilde\mu$ and the strain-rate tensor $\dot{\bm{\epsilon}} = (\nabla\bm{v} + \tra{\nabla\bm{v}})/2$, we can obtain from \cref{eq:archetype} the Navier-Stokes equations of motion. Indeed the variation of
$$
    \delta\tilde\rayleigh = \int_{\bar{V}} d\bar{V} \tilde{\bm{\sigma}} \cdot\delta\bm{\nabla v} = -\int_{\bar{V}} d\bar{V}\ \bm{v} \cdot \bm{\nabla}\cdot \tilde{\bm{\sigma}}
$$
with the deviatoric stress $\tilde{\bm{\sigma}}$ completes the right-hand sides of Euler equations~\cref{eq:eul_eul,eq:eul_lag}, e.g. in the Lagrangian picture of the latter
\begin{equation}\label{eq:nav_lag}
    \rho \left( \partial_t \bm{v}_\Lag + \bm{v}_\Lag\cdot\bm{\nabla v}_\Lag \right) + \bm{\nabla} p = \bm{\nabla}\cdot\tilde{\bm{\sigma}}.
\end{equation}
The fluctuations can be introduced into the Navier-Stokes \cref{eq:nav_lag} by including the stochastic terms of the Rayleighian \cref{eq:ray_memo_wave,eq:hydro}.

The potential $\mathcal{U}_\rho$ is the anti-derivative of $p/\rho$, where the pressure $p$ is given by an equation of state, as typically used to solve the Navier-Stokes equations. In the case of incompressible fluids, the pressure is introduced through a Lagrange multiplier $\eta$ to the constraint of constant mass density $\rho = \bar{\rho}$, by replacing $\mathcal{U}_\rho$ with
\begin{equation}\label{eq:eta}
    \bar{\mathcal{U}} = \eta (\rho - \bar\rho),
\end{equation}
which leads to $p = \eta \rho$. From~\cref{eq:drho_dt} we then obtain the usual condition imposed on incompressible fluids
\begin{equation}\label{eq:incompressible}
    \frac{d\rho}{dt} =- \rho \nabla\cdot \bm{v} = 0.
\end{equation}

\subsection{Microscopic connection}
The connection between molecular and macroscopic observables has both practical and theoretical value, since it provides a consistent definition of numerically measurable, fluctuating thermodynamic observables in atomistic simulations~\cite[Sec. 3]{evans2007statistical}, and analytical tools for statistical mechanics~\cite{Dean1996}. However, the representation of a microscopic stress poses conceptual difficulties that are still revisited in theoretical publications~\cite{Shi2023,Yuan2025,Lewis2026,pb2026}.

Here we also revisit the definitions of microscopic stress with focus on its connection to the macroscopic action principle. Our analysis identifies variables $\bm{u}$ and $\partial_t \bm{u}$ with the collective barycentric coordinate and streaming velocity in fluids, respectively, thus generalizing the simplified treatment of \cref{app:lim} for solids.

For simplicity we consider a single-component fluid with $N$ molecules of mass $m$ and pairwise interactions $e\big(\bm{q}(i)-\bm{q}(j)\big)$ between particles $i,j=1,2,...,N$ at positions $\bm{q}(i)$ and $\bm{q}(j\ne i)$ in a Cartesian unit cell of volume $\cell$ under \textit{periodic boundary conditions}. The total microscopic action can then be cast as a multiple of the action in the unit cell centered at a point $\bar{\bm{B}}$:
\begin{equation}\label{eq:actioncg}
    \mathcal{S}_\mic \propto \int dt \mathcal{L}_\mic
\end{equation}
with the Lagrangian
$$
    \mathcal{L}_\mic(\bar{\bm{B}}) = \frac{1}{\cell}\sum_{i=1}^N \frac{m \dot{\bm{q}}(i)^2}{2} - \frac{U_\mic}{\cell}
$$
and the potential energy
$$
    U_\mic = \frac{1}{2} \sum_{i,j \ne i} e\big(\bm{q}(i)-\bm{q}(j)\big).
$$

To derive the microscopic representation of fluid dynamics for $\bm{v} = \partial_t \bm{u}$, we introduce a single-particle mass density $\rho_i(\bm{q}) = m \delta\big(\bm{q} -\bm{q}(i)\big)$, where $\bm{q} = \bar{\bm{B}} + \bm{r}$ is an arbitrary geometric point of the unit cell parameterized by $\bm{r}$. 
Then we define the macroscopic density
\begin{equation}\label{eq:mass_density}
    \rho(\bar{\bm{B}}) = \int_\cell \frac{d\cell(\bm{r})}{\cell} \sum_i \rho_i(\bar{\bm{B}} + \bm{r})= \frac{m N}{\cell} = \frac{\mass}{\cell},
\end{equation}
and the displacement field
$$
    \bm{u}(\bar{\bm{B}}) = \int_\cell \frac{d\cell(\bm{r})}{\mass}\sum_i\bm{r} \bm{\rho}_i(\bar{\bm{B}} + \bm{r}\big) - \bar{\bm{B}} = \bm{C}(\bar{\bm{B}}) - \bar{\bm{B}}
$$
of the center of mass $\bm{C}=N^{-1} \sum_i \bm{q}(i)$.

Note that boundary crossing events (b.c.) cause abrupt changes of $\bm{C}$, when a particle leaving the unit cell is replaced by its periodic image. These jump changes can be modeled as stochastic events related to the entropic processes of mass redistribution (\cref{app:mass}). However, the periodic boundary conditions ensure conservation of mass and allows us to focus on modeling the dynamics of the fluid streaming velocity
$$
    \bm{v}(\bar{\bm{B}}) = \partial_t \bm{C}(\bar{\bm{B}}) = 
    N^{-1} \sum_i \dot{\bm{q}}(i).
$$

By taking the time derivative of the linear momentum density in the Cartesian coordinates $a,b\in\{x,y,z\}$ and using Einstein notation,
\begin{equation}\label{eq:mic}
    \rho \partial_t v^a = \bar\nabla_b \sigma^{ba} - \rho v_b \bar\nabla^b v^a,
\end{equation}
we identify the total stress (\cref{app:mic})
\begin{equation}\label{eq:sigma}
    \sigma^{ab}(\bar{\bm{B}}) =  \frac{1}{\cell} \sum_i \left(
        m \dot{r}^a(i) \dot{r}^b(i)
        - \frac{\partial U_\mic}{\partial q^a(i)} q^b(i)
    \right),
\end{equation}
where $\dot{\bm{r}}(i) = \dot{\bm{q}}(i) - \bm{v}(\bar{\bm{B}})$ are the particles' \textit{peculiar} velocities,\footnote{i.e. with respect to the streaming velocity of the fluid} and $\bar\nabla_b = \partial / \partial\bar{B}^b$.

To illustrate how the microscopic action principle is connected to its macroscopic counterpart we exploit the formalism of a virtual variation $\delta\bm{u}$. Consider a perturbation 
\begin{equation}\label{eq:virtual_u}
    \bm{u}(t) + \delta\bm{u}(t)
    = \frac{1}{N} \sum_i \big(\bm{q}(i) + \delta \bm{q}(i)\big) - \bar{\bm{B}},
\end{equation} which uniformly displaces the position of each particle $\delta\bm{q}(i) = \delta\bm{u}(t)$. Then the variation of \cref{eq:actioncg} yields
\begin{multline}
    \frac{\delta\action}{\delta u^a} = \sum_i \frac{\delta\action}{\delta q^b(i)} \frac{\delta q^b(i)}{\delta u^b(t)} \propto - \frac{1}{\cell} \sum_i \left(
        m \ddot{q}_a(i) + \frac{\partial U_\mic}{\partial q_a(i)}
    \right) \\= - \rho \left(\partial_t v^a - \bar\nabla_b \sigma_{ba}\right) = 0,
\end{multline}
which entails \cref{eq:sigma}. By a similar token we also obtain \cref{eq:incompressible} from \eqref{eq:mass_density}:
\begin{equation}\label{eq:delta_rho}
    \delta \rho = -\int_\cell \frac{d\cell(\bm{r})}{\cell}\sum_i \delta q^b(i) \nabla_b \rho_i = -\rho \bar\nabla_b \delta u^b.
\end{equation}

The microscopic \cref{eq:mic,eq:sigma} combine terms, that constitute the conservative part of the macroscopic dynamics depending on $\rho$, $\bm{u}$, and $\bm{v}$, with dissipative and fluctuating contributions that are modeled as the external forces (\cref{sec:flu}). These external forces emerge from the \textit{peculiar} velocities of particles, which are typically assumed to represent the thermal motion of molecules, and the ensuing redistribution of particles' \textit{peculiar} positions $\delta \bm{r}(i) = \delta\bm{q}(i) - \delta\bm{u}(t)$, not captured by the imposed uniform virtual displacement \cref{eq:virtual_u} that enforces $\delta \bm{r}(i) \equiv 0$.

In the thermodynamic limit $N\to\infty$, under the assumptions of large-deviation theory~\cite{Touchette2009}, the macroscopic action emerges as $\action = N I$, where the large-deviation functional $I = \lim_{N\to \infty} \action_\mic / N$ derives from the microscopic action. Thanks to the identity $\sum_i \bm{r}(i) \equiv 0$, as the kinetic energy density we can pose
$$
    \frac{m}{2 V} \sum_i [\bm{v} + \dot{\bm{r}}(i)]^2 \asymp \frac{\rho}{2} \bm{v}^2 + \frac{3 N \kB T}{2 V},
$$
in which the second term is constant for the isothermal flow. As the potential energy under the periodic boundary condition we must impose the incompressibility condition~\cref{eq:incompressible}, which introduces pressure through a Lagrange multiplier (\cref{sec:hydro}) and \cref{eq:delta_rho}. The discrepancies between \cref{eq:sigma} and the Euler \cref{eq:eul_lag} should then be interpreted as the dissipative part of the Navier-Stokes~\cref{eq:nav_lag} with optional \citet{LandauLifshitz1959Fluctuations} fluctuating terms [cf. \crefrange{eq:ray_memo_wave}{eq:hydro} in \cref{app:fdt_hooke}].

\section{\label{sec:num}Numerical examples}

\begin{figure*}[ht!]
    \includegraphics[width=2\columnwidth]{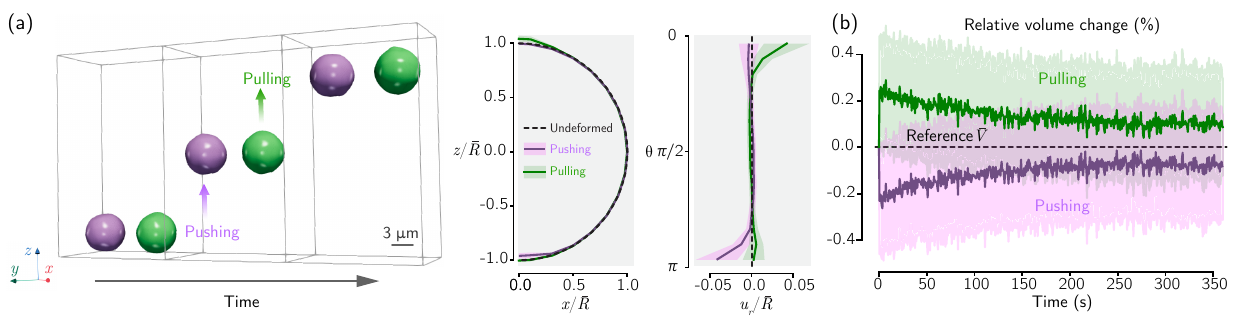}
    \caption{\label{fig:num}Numerical simulations of nuclei subject to external point-source forces and spontaneous fluctuations. (a) Simulations, starting from the initial configuration of a perfect ball (Video~\ref{vid:num}), describe the evolution of nuclear shape for a pushing (purple) and pulling (green) force of magnitude $z_\ext$ applied respectively at the south or north pole. Snapshots at representative frames \SI{66}{s}, \SI{126}{s}, \SI{180}{s} (left panel), and average shapes at \SI{180}{s} over 100 realizations and the azimuthal angle show a dent at the bottom under the pushing protocol and a bulge at the top when pulled. (b) Nuclear volume dynamics across 100 realizations indicate their tendency to compress or extend relative to the reference value $\bar{V}$ in response to pushing and pulling, respectively (confidence intervals: one standard deviation). The parameters of the simulations~\SM{Sec.~\smParams} correspond to realistic biophysical estimates (reference nucleus radius $\bar{R} = \SI{3}{\micro\meter}$, dissipative constant $\bar\gamma = \SI{10}{\kilo\gram\per\meter\squared\per\micro\second}$, Young modulus $E = \SI{5}{\kilo\pascal}$, Poisson ratio $\nu = 0.4$, fluctuation amplitude $T_{\rm eff} = \SI{40}{\atto\joule/\kB}$, external-force magnitude $z_\ext = \SI{1.2}{\nano\newton}$.)}
\end{figure*}

\begin{video}
    \setfloatlink{https://github.com/rbelousov/hookeanity/blob/main/sm/video_s1.mp4}
    \caption{\label{vid:num}Numerical simulations of nuclei subject to external point-source forces (green: pushing, magenta: pulling) and spontaneous fluctuations.
    }
\end{video}

To illustrate our theory by a numerical example for a representative system of biological interest, we derive a quantitative description of the shape dynamics of cell nuclei using experimental measurements from live-imaging and biophysical assays~\cite{Shin1999,Guilak2000,Caille2002,Dahl2004,Dahl2005,GLADILIN2010,Stephens2017,Schller2022,Estabrook2021,Jackson2023}. In particular, we simulate how nuclei respond to deterministic and stochastic cytoskeletal forces acting on their outer surface. Such directed cytoskeletal forces control nuclear positioning and movement within the cell, for example during interkinetic nuclear migration~\cite{MaiaGil2024,Spear2012}. Specifically, we consider the overdamped \cref{eq:overdamped} without active terms for an elastic object of a spherical reference shape of radius $\bar{R}$ in the form~\SM{Sec.~\smNum}
\begin{equation}\label{eq:sim}
    \frac{\delta(\xi-1)}{\bar{R}\xi^2}\bar{\gamma} \partial_t\bm{u} = \bm{f}_\ext(t) -  \bm{\nabla}\cdot\bm{\sigma}_\ela + \bm{f}_\flu(t).
\end{equation}

Note that second-order effects of finite rotations of a bounded Hookean body require the decomposition of the body motion into translational $\bm A(t)$ and rotational $\bm{\mathsf{R}}(t)$ components with respect to the centroid of the reference configuration (\cref{app:rot}):
\begin{equation}\label{eq:lab}
    \bm B = \bm A  + \bm{\mathsf{R}} \cdot \big(\bar{\bm{B}}(\xi,\theta,\phi) + \bm{u}(t, \xi, \theta, \phi)\big),
\end{equation}
where $\bm{B}$ and $\bm{u}$ represent, respectively, the configuration in the \textit{laboratory coordinates}\footnote{We assume that the centroid of the reference configuration $\bar{\bm{B}}$ is at the origin.} and a pure deformation in the \textit{internal reference frame}. Since the stress is linear in $\bm{u}$, the bulk degrees of freedom can be integrated out explicitly. The dynamics is therefore determined by the evolution of the surface \SM{Eq.~\eqSurf}
\begin{equation}\label{eq:flat}
    \bar{\gamma} \partial_t\bm{u}|_{\xi=1} = -\bm{\bar{n}}\cdot\bm{\sigma}_\ela|_{\xi=1} + \sqrt{2\kB T \bar{\gamma}} \partial\bm{w} + \bm{f}_\ext|_{\xi=1},
\end{equation}
with the outward normal $\bar{\bm{n}}$ given by \cref{eq:normal}.

By using representative values of the relevant physical parameters \SM{Sec.~\smParams}, we solve \cref{eq:flat} numerically~\cite{code} for the displacement field $\bm{u}$ expanded in a hybrid basis set of pseudospectral and radial-basis functions~\cite{supp,Flyer2009,Wright2010} with respect to the internal reference frame~\SM{Sec.~\smNum}. As the external driving we consider a point-source force $\bm{f}_\ext = -\bm{\nabla}\cdot\bm{\sigma}_\ext$ of magnitude $z_\ext$ acting on the body surface upward at the north or south pole (pushing or pulling) as viewed from the laboratory coordinates. 

Figure~\ref{fig:num} illustrates the effect of pushing and pulling forces acting on the nucleus along a specific axis, motivated by the example of interkinetic nuclear migration within the highly elongated cells of pseudostratified epithelia~\cite{MaiaGil2024}. This process arises in a variety of tissues and can involve both mechanisms of propulsion~\cite{Norden2017,Strzyz2016,Spear2012}. For simplicity, we model the active cytoskeletal fluctuations in the nuclear environment with an \textit{effective} temperature $T_{\rm eff} = 1 \times 10^4\ T_0$~\cite{Cugliandolo1997,Polettini2019}, though note that in principle the framework permits any fluctuation model, including memory or feedback effects. We find that pushing or pulling forces, comparable in magnitude to the force generated by approximately 300 molecular motors, lead to similar velocities but distinct deformations~(Video~\ref{vid:num}): pulling elongates the sphere into a prolate spheroid, whereas pushing renders its shape oblate [\cref{fig:num}(a)].

Interestingly, the volume of cell nuclei reflect the kind of driving protocol applied [\cref{fig:num}(b)]. When pushed or pulled, they respectively compress or expand relative to the reference configuration. This property provides an example where qualitative features of mechanical driving may be deduced from volume measurements alone.

\section{\label{sec:disc}Discussion}
We present a unified theory of fluctuating active soft matter, connecting microscopic descriptions of particle-level processes with the continuum-scale dynamics of finite bodies for both elastic and viscous media. We discuss examples, including Hookean elastodynamics, which permit an exact formal solution to expose the underlying structure of the theory outlined in \cref{sec:action}. This same structure 
can also be identified in more general contexts or extended to incorporate other physical phenomena. Here we briefly discuss such extensions. 

\paragraph*{General potentials}.---Besides the hydrodynamic potential (\cref{sec:hydro}), one could consider other generalizations, e.g. the family of soft-droplet potentials
\begin{align}\label{eq:droplet}
    U_\soft =& \Phi_\soft + \Psi_\soft,\\
    \Phi_\soft =& \gamma_\soft \int_{\partial V} dS
        \nonumber\\
        =& \gamma_\soft \iint d\theta d\phi\ |\partial_\theta \bm{B}(\bm{u}) \times \partial_\phi\bm{B}(\bm{u})|,\\\label{eq:volume_soft}
    \Psi_\soft =&\frac{\kappa_\soft}{2}\left(V - V_\soft\right)^2
        \nonumber\\
        =& \frac{\kappa_\soft}{2}\left(\iiint d\xi d\theta d\phi \sqrt{g(\nabla\bm{u})} - V_\soft\right)^2
\end{align}
with the surface and volume contributions $\Phi_\soft$ and $\Psi_\soft$, in which the constants $\gamma_\soft$, $\kappa_\soft$, and $V_\soft$ stand for the surface tension, volume elasticity, and preferred volume of the droplet, respectively. Instead of $\Phi_\soft$ one may also use the Helfrich potential~\cite{Salbreux2017Helfrich,Lipowsky1991Helfrich,Lipowsky2021Helfrich,Ulrich2019}
\begin{equation}\label{eq:helfrich}
    \Phi_\helfrich = \int_{\partial{\bar{V}}} dS\ \left[
        \frac{\kappa_{\rm B}}{2} (\tr\bm{\mathcal{C}}(\bm{u}) - H_0)^2 + \kappa_{\rm G} \frac{\mathcal{C}(\bm{u})}{g(\bm{u})}
    \right],
\end{equation}
in which the constants $\kappa_{\rm B}$, $\kappa_{\rm G}$, and $H_0$ are the bending rigidity, Gaussian modulus, and spontaneous curvature, whereas $\mathcal{C}_{ij} = \partial_i\hat{\bm{n}}(\bm{u}) \cdot \partial_j\bm{B}(\bm{u}) = - \hat{\bm{n}}(\bm{u}) \cdot \partial_i\partial_j\bm{B}(\bm{u})$ and $\mathcal{C}(\bm{u})$ are respectively the extrinsic curvature tensor\footnote{In soft matter, the term curvature tensor (or shape operator) typically refers to the second fundamental form, which should not be confused with the Riemann curvature tensor of intrinsic differential geometry.} and its determinant evaluated with the unit normal $\hat{\bm{n}}$ of the deformed surface. Or the potential of a thin elastic shell of depth $h$ \cite[\S11--15]{landau2005theory},
\begin{multline}\label{eq:thin_shell}
    \Phi_\shell = \int_{\partial{\bar{V}}} dS\left[
        \frac{\bar{\lambda}_\shell}{2}(\tr{\bm{\epsilon}_{\theta:\phi})^2}
        + \bar{\mu}_\shell \fr{\bm{\epsilon}_{\theta:\phi}}^2
        \right.\\\left.
        + \frac{h^2}{24} \left(
            \bar{\lambda}_\shell (\tr{\bm{\Delta \mathcal{C}}})^2
            + 2 \bar{\mu}_\shell \fr{\bm{\Delta \mathcal{C}}}^2
        \right)
    \right],
\end{multline}
with the two-dimensional surface strain $\bm{\epsilon}_{\theta:\phi}$ and the Lam\'e coefficients $\lambda_\shell$, $\mu_\shell$, and a curvature change $\Delta \mathcal{C}_{ij} = \mathcal{C}_{ij} - \bar{\mathcal{C}}_{ij}$ with respect to the reference configuration.

All the potentials defined by \crefrange{eq:droplet}{eq:thin_shell} are functionals of the displacements $\bm{u}$ included into the metric $\bm{g}(\bm{u})$, the curvature $\bm{\mathcal{C}}$, or the surface strain $\bm{\epsilon}$. All these functionals lead to nonlinear expressions for the stress and traction forces arising from $\delta U_\soft / \delta\bm{u}$ as defined by \cref{eq:stress}. Moreover \cref{eq:volume_soft} introduces nonlocal interactions.

\paragraph*{Diffusive flows}.---While the diffusive processes in solids are typically negligible over the hydrodynamic time scales, in fluids redistribution of mass may play an important role. Such phenomena represent entropic processes described by the principles of statistical mechanics~\cite{Dean1996,Bertini2015,Belousov2018,Wrthner2023,Belousov2026} that provide specific constitutive relations for the mass fluxes. Such systems are then described by coupled transport equations for momentum and mass density \cref{eq:mass_continuity}. Moreover, the continuity equation for momentum acquires a contribution in the total time derivative of \cref{eq:time}, e.g. the Cauchy~\cref{eq:cauchy} must be modified to include a correction from \cref{eq:drho_dt}:
\begin{equation}\label{eq:diffusive}
    \bar\rho\partial_t^2\bm{u} + \partial_t\bar\rho\ \partial_t\bm{u} - \bm{\nabla}\cdot\bm{\sigma}_\ela = \bm{0},
\end{equation}
and must be supplemented by the appropriate mass continuity~\cref{eq:mass_continuity}. Note that the above equation refers to the Eulerian reference frame as considered in \cref{app:mass}. In the Lagrangian picture one needs to take into consideration also the time dependence of $\bar{\bm{g}}$.

\paragraph*{Conclusion}.---The action principle represents an integrative framework for unifying diverse domain-specific physical models of elastic and viscous soft matter. Indeed, many of these models can be approximated by Hookean elastodynamics via power-series expansions, e.g. to inspect or compare the elastic linear response.

From the stationary action principle, through the Hamiltonian formalism and the fluctuation-dissipation theorem we obtain a consistent thermodynamic formalism. Its thermodynamic consistency relies on the foundations already well established for classical mechanical systems~\cite{Seifert2025,Tuckerman2023,Peliti2021}.

Beyond offering a foundation for comparing different approaches, our theory provides a means of translating quantitative data between models, e.g. relating parameters of nonlinear and composite models~\crefrange{eq:droplet}{eq:thin_shell} to the effective moduli of the linearized elasticity. Such a means is particularly timely, as advancing imaging techniques are enabling increasingly powerful methods for physical analysis and inference~\cite{Jackson2023,MaiaGil2024,Jurado2025,Drozdowski2026}, allowing information to be extracted from experiments that are not amenable to direct measurement.

Our modeling example demonstrates how nuclear shape reflects nuanced features of the forces exerted by its surroundings. Thus, the numerical methods used in our simulations may also serve as a basis for designing inference techniques, a direction that remains to be explored in future work.

\begin{acknowledgments}
V.B., M.L., I.D.E., L.W., I.R.G., R.B. and A.E. acknowledge funding from the EMBL. This work was supported by a Research Grant from HFSP (Ref.-No: RGP007/2023). L.R. gratefully acknowledges support from the Italian Ministry of University and Research (MUR) through the grant PRIN2022-PNRR project (No. P2022Z7ZAJ) ``A Unitary Mathematical Framework for Modelling Muscular Dystrophies'' (CUP: E53D23018070001) funded by the European Union - Next Generation EU. The authors are grateful to Jenna Elliott, Patrick Jentsch, Samuele Massimi, Charalampos Pozoukidis for their feedback.

The authors also acknowledge the use of AI tools~\cite{chatgpt,codex} for bibliographic search, as well as for language and coding assistance. All references were verified. The generated code was reviewed, manually refined, commented, and documented. The authors take full responsibility for the scientific content, interpretation, and conclusions of the manuscript.
\end{acknowledgments}

\section*{Data Availability Statement}
The data that support the findings of this study were generated by numerical simulations. The source code and parameters used to generate the simulations is publicly available~\cite{supp,code}.

\appendix
\section*{Appendix}
\subsection{\label{app:osc}Harmonic oscillator}
The condition of least action for the harmonic oscillator introduced by \cref{eq:action,eq:lag_osc} yields the Euler-Lagrange equations~\cite[Chapter 2]{Goldstein2012}
\begin{align}\label{eq:dS_osc_first}
    \frac{\delta\action_{1:2}}{\delta X_1} = - \frac{d}{dt} \left(M_1 \dot{X}_1\right) - K (X_2 - X_1 - L) &= 0,\\\label{eq:dS_osc_second}
    \frac{\delta\action_{1:2}}{\delta X_2} = - \frac{d}{dt} \left(M_2 \dot{X}_2\right) + K (X_2 - X_1 - L) &= 0,
\end{align}
which are equivalent to the Newton equations of motion
\begin{equation}\label{eq:newt_osc}
    M_i \ddot{X}_i = - \frac{\partial U_{1:2}(X_1, X_2)}{\partial X_i},
\end{equation}
with the potential
\begin{equation}\label{eq:pot_osc}
    U_{1:2} = \frac{K}{2} (X_2 - X_1 - L)^2.
\end{equation}

By defining the canonical momentum
\begin{equation}\label{eq:mom_osc}
    p_i = \frac{\partial \mathcal{L}_{1:2}}{\partial\dot{X}_i} = M_i \dot{X}_i
\end{equation}
and applying the Legendre transformation, we obtain the Hamiltonian
\begin{equation}\label{eq:ham_osc}
\mathcal{H}_{1:2} = \sum_i p_i \dot{X}_i - \mathcal{L}_{1:2} = \sum_i \frac{M_i \dot{X}_i^2}{2} + U_{1:2},
\end{equation}
conserved by \cref{eq:newt_osc}:
\begin{equation}\label{eq:conservation_osc}
    \frac{d\mathcal{H}_{1:2}}{dt} = \sum_i \dot{X}_i \left[
        M_i \ddot{X}_i + \frac{\partial U_{1:2}}{\partial X_i}
    \right] \equiv 0.
\end{equation}

\subsection{\label{app:lim}Elasticity in one dimension}
The Hookean theory of elasticity can be related to the continuum limit of a chain of harmonic oscillators~\cite[Sec.~1.2]{kardar2007statistical}. Here we briefly summarize this relation in one dimension by considering an extension of Eq.~\eqref{eq:lag_osc} for $N$ particles connected by $N$ springs of stiffness constants $K_{i=1,2...N}$:
\begin{equation}\label{eq:lag_osc_N}
    \mathcal{L}_{1:N} = \sum_{i=1}^N \Big[
        \frac{M_i}{2}\dot{X}_i^2 \\ - \frac{K_i}{2} (X_{i+1} - X_i - L)^2
    \Big],
\end{equation}
in which we imply the periodic boundary condition $X_{N+1} = X_1$.

By substituting $X_i = \bar{X}_i + \delta X_i$ and $\dot{X}_i = \delta \dot{X}_i$ into \cref{eq:lag_osc_N}, where $\bar{X}_i = (i - 1) L$ is a \textit{reference configuration} (cf. \cref{app:ref}), we get
\begin{equation}\label{eq:lag_osci}
    \mathcal{L}_{1:N} = \sum_{i=1}^N \left[\frac{M_i}{2} \delta\dot{X}_i^2 - U_i\right],
\end{equation}
with the particle potential
\begin{equation}\label{eq:pot_i}
    U_i =\frac{K_i}{4} (\delta X_{i+1} - \delta X_i)^2 + \frac{K_{i-1}}{4} (\delta X_i - \delta X_{i-1})^2.
\end{equation}
The additional factor of $1/2$ in the above equation ensures that each interaction term appears just once after the summation of \cref{eq:lag_osci}. Then, by treating $x = i dx$ as a continuum variable with $dx = L$, we expand
$$
    \delta X_{i\pm1} = \delta X(x \pm dx) = \delta X(x) \pm \partial_x \delta X(x) dx + O(dx^2),
$$
and $K_{i-1} = K_i + O(dx)$. Substituted into the particle potential, these expansions yield
\begin{equation}\label{eq:pot_x}
    U_i = U(x) = \frac{K_i}{2} dx^2 \big(\partial_x \delta X(x)\big)^2 + O(dx^3).
\end{equation}
The continuum limit of \cref{eq:lag_osci} leads to
\begin{multline}\label{eq:lim}
    \mathcal{L}_{1:N} \simeq \sum_{i=1}^N dx \left[\frac{M_i / dx}{2} \delta\dot{X}(x)^2 - \frac{K_i dx}{2} (\partial_x \delta X)\right]\\
    \underset{N\to\infty, dx\to0}{\to} \int dx \left[
        \frac{\bar{\rho} (\partial_t u)^2}{2} - \frac{\bar{\lambda}}{2} (\nabla u)^2
    \right],
\end{multline}
in which the summation operator $\sum_i dx \to \int dx$ turns into the integral, $M_i/dx \to \bar{\rho}(x)$ and $K_i dx\to\bar{\lambda}(x)$ become the mass density $\bar{\rho}$ of the reference configuration and the space-dependent Lam\'e coefficient $\bar{\lambda}$, whereas the derivatives of the particles' displacements $\partial_x \delta X \to \nabla u$ with their velocities $\delta\dot{X}$ yield the gradient of the displacement field $\nabla u$ and its time derivative $\partial_t u$.

In the rest of the work we assume uniform Lam\'e coefficients $\bar{\lambda} = \lambda \equiv \const$.

\subsection{\label{app:ref}Reference frames}

\paragraph*{Eulerian picture.}---When constructing a general isomorphism between points $\bm{r} = (\xi, \phi, \theta)$ restricted to a unit ball $\xi\in[0,1]$, $\theta\in[0,\pi]$, $\phi\in[0,2\pi)$ and the configurations $\bar{\bm{B}}(\bm{R})$, $\bm{B} = \bm{u}(\bm{r}) + \bar{\bm{B}}(\bm{r})$, we define an \textit{Eulerian} reference frame~\cite[Sec. 9.9]{Goldstein2012}, since neither $\bm{r}$ nor $\bar{\bm{B}}$ depend on time. This isomorphism may be highly nonlinear and more complex than illustrated in \cref{fig:ref}. For instance, in certain configurations, there always exist some rays emanating from any internal point---not just from the geometric center or center of mass---that, at specific angles $(\theta, \phi)$, intersect the bounding surface at two or more distinct locations. Then the radial parametrization by $R(\theta,\phi)$ like in \cref{fig:ref} would not be viable for such solid angles $(\theta, \phi)$.

However, our theory is not restricted to the special case of bodies that admit a radial parameterization of the bounding surface, and can even be extended beyond spherical topologies. Since linear elasticity relies on covariant derivatives of the displacement vector, our approach is applicable as long as the manifold configurations $\bm{B}$ are differentiable and each coordinate of its points can be charted as $B^i(\xi^1,\xi^2,\xi^3)$ by using three parameters $\xi^{j=1,2,3}$.

\paragraph*{Lagrangian picture.}---Sometimes a time-dependent parametrization $\bm{r}(t)$, is more convenient than the Eulerian picture. A care must then be taken in expanding the total time derivative operator $d/dt$ that appears in variation of the action \cref{eq:daction_ela}. In particular, we observe
\begin{equation}\label{eq:lag_vel}
    \frac{d}{dt}\partial_t u^j = \partial_t^2 u^j + \partial_t r^i \nabla_i \bm{u}^j,
\end{equation}
whereas in the Eulerian picture we have $\partial_t r^i\equiv 0$.

Importantly, because the observer may design an arbitrary protocol of updating the parametrization $\bm{r}(t)$, the velocity $\partial_t \bm{r}$ does not represent a \textit{physical law}, even if chosen to depend on the system's state variables $\big(\bm{u},\partial_t\bm{u})$. The effects of the time-dependent reference coordinates $\bm{r}(t)$ should therefore be taken into account in the Euler-Lagrange equations~\cref{eq:time} after the variation of the action functional has been applied. Specifically, if we choose $\partial_t r^i(t) = \partial_t u^i(t)$ we obtain a \textit{Lagrangian picture}~\cite[Sec. 9.9]{Goldstein2012} of hydrodynamics, in which \cref{eq:lag_vel} entails the material derivative~\cite{Wrthner2023}
\begin{equation}\label{eq:material_time}
    \frac{d}{dt} = \partial_t + \partial_t \bm{u} \cdot\bm{\nabla}.
\end{equation}

\subsection{\label{app:strain} Linear strain}
For a small displacement $\bm{u}$, the deformation of a body $\bm{B} = \bar{\bm{B}} + \bm{u}$ with respect to the reference configuration $\bar{\bm{B}}$ (\cref{fig:ref}) is encoded by the metric tensors $\bm{g}$ and $\bar{\bm{g}}$ associated with the three-dimensional manifolds $\bm{B}(\xi, \theta, \phi)$ and $\bar{\bm{B}}(\xi, \theta, \phi)$. First, we define the basis tangent vectors conjugate to the coordinates $i\in\{\xi,\theta,\phi\}$ on the two manifolds,
$$
    \bar{\bm{e}}_i = \partial_i \bar{\bm{B}},\quad
    \bm{e}_i = \partial_i \bm{B} = \bar{\bm{e}}_i + \partial_i \bm{u}.
$$
and then the metric tensors $\bar{g}_{ij} = \bar{\bm{e}}_i \cdot \bar{\bm{e}}_j$ and
\begin{equation}\label{eq:gij}
    g_{ij} = \bm{e}_i \cdot \bm{e}_j = \bar{g}_{ij}
        + \bar{\bm{e}}_i \cdot \partial_j \bm{u} + \bar{\bm{e}}_j\cdot \partial_i \bm{u} + \partial_i \bm{u} \cdot\partial_j\bm{u}.
\end{equation}
To simplify \cref{eq:gij}, we need a few formulae from the differential geometry:
\begin{align}\label{eq:Christoffel}
    \nabla_j \bar{\bm{e}}_k =& \partial_j \bar{\bm{e}}_k = \bar\Gamma^i_{jk} \bar{\bm{e}}_i,\\\label{eq:j_u}
    \partial_j \bm{u} =& \partial_j u^k \bar{\bm{e}}_k + u^k \bar\Gamma^i_{jk} \bar{\bm{e}}_i
    = \nabla_j u^i \bar{\bm{e}}_i = \nabla_j u_i \bar{\bm{e}}^i,\\
    \partial_i \bm{u} \cdot \partial_j \bm{u} =& (\nabla_i u^k \bar{\bm{e}}_k) \cdot (\nabla_j u^l \bar{\bm{e}}_l) = \nabla_i u^k \nabla_j u^l \bar{g}_{kl} \nonumber\\=& \bm{\nabla u} \cdot \tra{\bm{\nabla u}}.
\end{align}
where $\Gamma^i_{jk}$ are Christoffel symbols,\footnote{
    Equation~\eqref{eq:Christoffel} accounts for the fact that deformations in three dimensions do not have a tangential component if embedded in a four-dimensional manifold and, thus, is a special case of the general geometric definition of Christoffel symbols.
} and we use the Einstein summation notation for the repeated pairs of upper and lower indices, with the upper ones corresponding to the contravariant tensor components. Since $\bar{\bm{e}}_i \cdot \partial_j \bm{u} = \nabla_j u_i$, from \cref{eq:gij} we get
\begin{equation}\label{eq:g_u}
    \bm{g} = \bar{\bm{g}} + \bm{\nabla u} + \tra{\bm{\nabla u}} + \bm{\nabla u} \cdot \tra{\bm{\nabla u}}.
\end{equation}

The metric tensor describes how the displacement field $\bm{u}$ deforms the distances between any two points $\bar{\bm{B}}_2$ and $\bar{\bm{B}}_1$ within some small neighborhood $\Delta\bar{\bm{B}} = \bar{\bm{B}}_2 - \bar{\bm{B}}_1$. After the deformation the squared distance
$$
    \Delta\bar{\bm{B}}^2 = \bar{g}_{ij} \Delta \bar{B}^i \Delta \bar{B}^j,
$$
becomes
\begin{multline}\label{eq:neighborhood}
    \Delta{\bm{B}}^2 = g_{ij} \Delta B^i \Delta B^j
        \\= \Delta{\bar{\bm{B}}}^2 + (\nabla_i u_j + \nabla_j u_i) \Delta \bar{B}^i \Delta\bar{B}^j + O\big((\nabla u)^2\big),
\end{multline}
spelled out up to quadratic order in the components of $\nabla\bm{u}$. Then we can decompose the tensor $\nabla\bm{u} = \bm{\epsilon} + \bm{\omega}$ into the symmetric, $\epsilon_{ij} = \epsilon_{ji} $, and the antisymmetric, $\omega_{ij} = - \omega_{ji}$, parts:
\begin{equation}\label{eq:nabla_u}
    \bm{\epsilon} = \frac{\bm{\nabla u} + \tra{\bm{\nabla u}}}{2},\quad
    \bm{\omega} = \frac{\bm{\nabla u} - \tra{\bm{\nabla u}}}{2},
\end{equation}
to cast \cref{eq:neighborhood} as
$$
    \Delta\bm{B}^2 - \Delta\bar{\bm{B}}^2 = 2 \epsilon_{ij} \Delta \bar{B}^i \Delta\bar{B}^j + O\big((\nabla u)^2\big).
$$
Whereas the tensor of \textit{linear strain} $\bm{\epsilon}$ describes infinitesimal changes of distances in various directions of the neighborhood, the tensor $\bm{\omega}$ describes infinitesimal rotations of the whole neighborhood.

\subsection{\label{app:hooke}Hookean elastic potential}
In the curvilinear coordinates the Hookean elastic potential is constructed to be invariant under translations and rotations. This condition is fulfilled when the expression depends only on the linear strain $\bm{\epsilon}$ formed by symmetrizing the translation-invariant tensor of the displacement gradient $\bm{\nabla u}$ and, thus, removing its rotational part $\bm{\omega}$ (\cref{app:strain}).

The most general expression of the Hookean potential, which must be a coordinate-independent expression---quadratic in the displacement $\bm{u}$ and, hence, also in the linear strain $\bm{\epsilon}$,---takes the form~\cite[Chapter I]{landau2005theory}, \cite[Sec.~7.2]{irgens2008continuum}
$$
    \mathcal{U} = \frac{1}{2}K^{ijkl} \epsilon_{ij} \epsilon_{kl}
$$
with a fourth-order elastic-modulus tensor $K^{ijkl}$.
However, the symmetry of isotropic systems allows at most two independent scalars composed from the symmetric tensor components $\epsilon_{ij}$. Using the matrix representation of $\bm{\epsilon}$ we can form three invariants: trace, determinant, and the Frobenius norm. The determinant of a third-rank matrix is not convenient for a quadratic expression, because it would produce terms of cubic order. Therefore we are left with the trace,
\begin{equation}\label{eq:tr}
    \tr \bm{\epsilon} = \mathds{1} : \bm{\epsilon} = \epsilon^i_i = \bar{g}^{ij}\epsilon_{ij},
\end{equation}
and the Frobenius norm generalized for curvilinear coordinates 
\begin{equation}\label{eq:fr}
    \fr{\bm\epsilon} = \sqrt{\bm{\epsilon}:\bm{\epsilon}} = \sqrt{\epsilon^{ij} \epsilon_{ij}} = \sqrt{\bar{g}^{ik} \bar{g}^{jl} \epsilon_{ij} \epsilon_{kl}}.
\end{equation}
Here $\mathds{1}$ is a unit matrix, $\bar{g}^{ij}$ stand for the components of the inverse metric tensor, and the colon indicates double contraction.

By forming a linear combination of squared \cref{eq:tr,eq:fr} with constants $\bar\lambda$ and $\bar\mu$, we obtain the quadratic expression for the Hooke potential \cref{eq:hooke_ela}.

\subsection{\label{app:stress}Stress}
The standard definition of elastic stress for a Hookean body reads~\cite[Chapter I]{landau2005theory}
\begin{equation}\label{eq:stress_std}
    \bm{\sigma}_\ela  = \frac{\partial \mathcal{U}_{\ela}}{\partial \bm{\epsilon}} = \lambda \tr(\bm{\epsilon})\bar{\bm{g}} + 2 \mu \bm{\epsilon},
\end{equation}
which follows from \cref{eq:hooke_ela} and in components means
$$
    (\bm{\sigma}_\ela)_{mn} = \bar{g}_{mi} \bar{g}_{nj} \frac{\partial \mathcal{U}_\ela}{\partial \epsilon_{ij}}
    = \lambda \bar{g}_{mn} \bar{g}^{kl} \epsilon_{kl} + 2 \mu \epsilon_{mn}.
$$
In this work we use a more convenient definition applicable to strain-dependent potentials like $U_\ela[\bm{\epsilon}]$,
\begin{equation}\label{eq:stress}\bm{\sigma}_\ela = \frac{\delta U_\ela}{\delta \bm{\nabla u}} = \frac{\partial\mathcal{U}_\ela}{\partial \bm{\nabla u}},
\end{equation}
which is consistent with \cref{eq:stress_std} when expanded in components
\begin{multline}
     \bar{g}_{mi} \bar{g}_{nj}\frac{\partial U}{\partial \epsilon_{ab}} \frac{\partial\epsilon_{ab}}{\partial\nabla_i u_j}
     \\= \bar{g}_{mi} \bar{g}_{nj} (\bm{\sigma}_\ela)^{ab} \frac{\delta_a^i\delta_b^j + \delta_a^j\delta_b^i}{2} = (\bm{\sigma}_\ela)_{mn}.
\end{multline}
Here $\delta_i^a$ is the Kronecker delta.

\subsection{\label{app:eul_lag_hooke}Euler-Lagrange equations for a Hookean body}
The first variation of the action $\action$ expanded with \cref{eq:lag_ela,eq:hooke_ela,eq:stress} yields
\begin{equation}\label{eq:daction}
    \delta\action = \int_0^{\mathcal{T}} dt\, \int_{\bar{V}} d\bar{V}\, \left(
        \bar\rho \partial_t\bm{u} \cdot \partial_t\delta\bm{u}
        - \bm{\sigma}_\ela : \bm{\nabla} \delta \bm{u})
    \right],
\end{equation}
in which we used \cref{eq:stress}. Integration of the first term by parts with respect to time yields
\begin{multline}\label{eq:time}
    \int_0^{\mathcal{T}} dt\, \int_{\bar{V}} d\bar{V}\, 
        \bar\rho \partial_t\bm{u} \cdot \partial_t\delta\bm{u}
        \\= \cancel{\bar\rho \partial_t\bm{u} \cdot \delta\bm{u}\Big|_{t=0}}^{\mathcal{T}} - \int_0^{\mathcal{T}} dt \int_{\bar{V}} d\bar{V}\left(\frac{d}{dt} \big(\bar\rho \partial_t\bm{u}\big)\right) \cdot \delta\bm{u},
\end{multline}
where the boundary terms are canceled out, because the standard action principle requires that the variation of displacement vector at the initial and final times vanish, $\delta\bm{u}|_{t=0} = \delta\bm{u}|_{t={\mathcal{T}}}$.

Likewise, integration of the second term in the integrand of \cref{eq:daction} by parts over the volume gives
\begin{multline}\label{eq:space1}
    \int_{\bar{V}} d\bar{V}\,
        \bm{\sigma}_\ela : \bm{\nabla} \delta \bm{u}
    = \oint_{\partial \bar{V}} d\bm{\bar{S}} \cdot \bm{\sigma}_\ela \cdot \delta\bm{u}
    \\ - \int_{\bar{V}} d\bar{V} (\bm{\nabla} \cdot\bm\sigma_\ela) \cdot\delta\bm{u},
\end{multline}
where $d\bm{S}$ is the oriented element of the surface $\partial\bar{V}$ that bounds the volume $\bar{V}$ of the body.

The boundary surface term in \cref{eq:space1} suggests the zero traction force
\begin{equation}\label{eq:zero_traction}
    \hat{\bar{\bm{n}}} \cdot \bm{\sigma}_\ela = \bm{0},
\end{equation}
with the unit normal $\hat{\bar{\bm{n}}}$ of the reference configuration surface,
as \text{the natural boundary condition}, which reduces \cref{eq:space1} to
\begin{equation}\label{eq:space2}
    \int_{\bar{V}} d\bar{V}\,
        \bm{\sigma}_\ela : \bm{\nabla} \delta \bm{u}
    = - \int_{\bar{V}} d\bar{V} (\bm{\nabla} \cdot\bm\sigma_\ela) \cdot\delta\bm{u}.
\end{equation}
Equations~\eqref{eq:time} and \eqref{eq:space2}, substituted into \eqref{eq:daction}, give \eqref{eq:daction_ela}.

When the boundary condition \cref{eq:zero_traction} is not applicable, the boundary surface term in \cref{eq:space1} can be transformed into a volume integral
\begin{equation}\label{eq:space3}
    \oint_{\partial \bar{V}} d\bm{\bar{S}} \cdot \bm{\sigma}_\ela \cdot \delta\bm{u}
        = \int_{\bar{V}} d\bar{V}\, \frac{\delta(\xi - 1)}{\sqrt{\bar{g}}} \bar{\bm{n}} \cdot \bm{\sigma}_\ela \cdot \delta\bm{u},
\end{equation}
with help of the Dirac delta function and the outward normal (non-unit) vector
\begin{equation}\label{eq:normal}
    \bar{\bm{n}} = \bar{\bm{e}}_\theta \times \bar{\bm{e}}_\phi.
\end{equation}
Equation~\eqref{eq:space3} introduces then a source term into the Cauchy equations of motion~\eqref{eq:cauchy}:
\begin{equation}\label{eq:dirac_delta}
    \bar\rho\partial_t^2 u - \bm{\nabla}\cdot\bm{\sigma}_\ela = - \frac{\delta(\xi - 1)}{\sqrt{\bar{g}}} \bar{\bm{n}} \cdot \bm{\sigma}_\ela|_{\xi = 1},
\end{equation}
where $\bm{\sigma}_\ela(\xi,\theta,\phi)|_{\xi = 1}$ must be specified as the boundary condition on $\bm{u}$ over $\partial \bar{V}$:
$$
    \bm{\sigma}_\ela(\bm{\nabla u}) = \bm{\sigma}_\ela(1,\theta,\phi).
$$

\subsection{\label{app:mass}Mass conservation and continuity equation}
In general, the total body mass
\begin{equation}\label{eq:mass}
    \mass = \int_{\bar{V}} d\bar{V} \bar{\rho} = \int_V dV \rho,
\end{equation}
is conserved under the deformation of the volume elements (\cref{fig:ref})
\begin{equation}\label{eq:dV}
    d\bar{V} = \sqrt{\bar{g}} d\xi d\theta d\phi \;\to\;
    dV = \sqrt{g} d\xi d\theta d\phi,
\end{equation}
where $\bar{g}$ and $g$ are the determinants of the metric tensors associated with the three-dimensional manifolds $\bar{\bm{B}}$ and $\bm{B}$, respectively. From \cref{eq:mass} we therefore get
\begin{equation}\label{eq:mass_conservation}
    \Delta \mass = \int_0^1 d\xi \int_0^\pi d\theta \int_0^{2\pi} d\phi\, (\sqrt{g} \rho - \sqrt{\bar{g}} \bar{\rho}) = 0.
\end{equation}
In particular, this equality holds trivially when the integrand vanishes, as in the ideal elastic solid~\cite[Sec.~5.5.2]{irgens2008continuum} where $\rho dV = \bar{\rho} d\bar{V}$ entails
\begin{equation}\label{eq:rho}
    \rho = \frac{\sqrt{\bar{g}}}{\sqrt{g}} \bar{\rho}.
\end{equation}
The above equation describes how the mass density $\rho$ changes as a result of pure deformation, when the mass in the reference configuration is not redistributed and the density $\bar\rho$ remains fixed, i.e.\ $\partial_t \bar{\rho} = 0$.

In the more general case the density $\bar\rho(t)$ may be an explicit function of time. Then from the last equality of \cref{eq:mass} we have
\begin{equation}\label{eq:dM}
    \frac{d}{dt} \mass = \iiint d\xi d\theta d\phi\, \left(\frac{d\sqrt{g}}{dt} \rho + \sqrt{g} \frac{d\rho}{dt}\right) = 0,
\end{equation}
in which
\begin{multline}\label{eq:dg_dt}
    \frac{d \sqrt{g}}{dt} = \frac{1}{2\sqrt{g}}\frac{d g}{dt}
        = \frac{\sqrt{g}}{2} g^{ij} \frac{d g_{ij}}{dt}\\
        = \frac{\sqrt{g}}{2} g^{ij} \partial_t g_{ij} + \frac{\sqrt{g}}{2} g^{ij} \partial_t(\nabla_i u_j + \nabla_j u_i)
        \\= \sqrt{g} g^{ij} \nabla_i \partial_t u_j
        = \sqrt{g} \bm{\nabla} \cdot \partial_t \bm{u},
\end{multline}
where we use the Jacobi formula for the determinant, then \cref{eq:g_u}, the symmetry of the metric tensor $g_{ij} = g_{ji}$ and its independence of time in the Eulerian picture. From \cref{eq:rho} we further get
\begin{multline}\label{eq:drho_dt}
    \frac{d \rho}{dt} = \frac{\bar\rho}{\sqrt{g}} \frac{d \sqrt{\bar{g}}}{dt} + \sqrt{\bar{g}} \bar\rho \frac{d \left(g^{-1/2}\right)}{dt} + \frac{\sqrt{\bar{g}}}{\sqrt{g}}\partial_t \bar{\rho}
        \\
        = - \frac{\sqrt{\bar{g}}}{\sqrt{g}} \bar\rho \bm{\nabla}\cdot\partial_t\bm{u} + \frac{\sqrt{\bar{g}}}{\sqrt{g}} \partial_t \bar\rho
        = - \rho \bm{\nabla}\cdot\partial_t\bm{u} + \partial_t \rho.
\end{multline}
By substituting \cref{eq:dg_dt,eq:drho_dt} into \eqref{eq:dM} we obtain
\begin{equation}\label{eq:mass_closed}
    \frac{d\mass}{dt} = \int_V dV \partial_t \rho = \int_{\bar{V}} d\bar{V} \partial_t \bar{\rho} = 0.
\end{equation}

The scalar $\partial_t \rho$, whose integral \cref{eq:mass_closed} over a closed volume $V$ vanishes, can be expressed as a covariant divergence of a vector---the negative of a diffusion current~$\bm{j}$:
\begin{equation}\label{eq:mass_continuity}
    \partial_t \rho = \frac{\sqrt{\bar{g}}}{\sqrt{g}}\partial_t\bar\rho= - \bm{\nabla} \cdot \bm{j},
\end{equation}
which represents the \textit{mass continuity equation}. The explicit form of the diffusive flux $\bm{j}$, which has an entropic origin, should be based on the theory of statistical mechanics, e.g.\ the Fick's law, the Kawasaki-Dean equation~\cite{Dean1996}, or the variational principles~\cite{Bertini2015,Belousov2026}.

\subsection{\label{app:act}Active stress}
To show that the active-stress term in \cref{eq:pot_act} generalizes the analogous term in \cref{eq:pot_t_osc}, first consider the harmonic-chain model from \cref{app:lim}. By introducing $K_i(t)$ and $F_i(t)$ into the potential terms \cref{eq:pot_i}, we get
\begin{multline}\label{eq:pot_i_t}
    U_i(t) = \frac{K_i}{4} (\delta X_{i+1} - \delta X_i)^2 + \frac{K_{i-1}}{4} (\delta X_i - \delta X_{i-1})^2
    \\ - \frac{F_i}{2} (\delta X_{i+1} - \delta X_i) - \frac{F_{i-1}}{2}(\delta X_i - \delta X_{i-1}),
\end{multline}
which in \cref{eq:pot_x} implies
\begin{multline}\label{eq:pot_x_t}
    U_i(t) = U(t, x) = \frac{K_i}{2} dx^2 \big(\partial_x \delta X(x)\big)^2 \\
    - F dx \partial_x\delta X(x) + \frac{dx^2}{2}\partial_x F \delta X(x)
     + O(dx^3).
\end{multline}
Because \cref{eq:pot_i_t} does not change the kinetic part of the Lagrangian \eqref{eq:lag_osci}, in the continuum-limit \cref{eq:lim} we only change the potential part
\begin{multline}\label{eq:lim_act}
    \sum_i dx \left[
        \frac{K_i(t) dx}{2} \big(\partial_x \delta X(x)\big)^2
        \right.\\\left.
        - F \partial_x\delta X(x) + \frac{dx}{2}\partial_x F \delta X(x)
    \right] \\\to \int dx \left[
        \bar\lambda_t (\nabla u)^2 - f_0(t) \nabla u + u \nabla f_1(t)
    \right],
\end{multline}
in which we pose $F \to f_0$ and $dx \partial_x F / 2 \to \nabla f_1$. However, note that the last term of the right hand side we can integrate by parts to get
\begin{multline}
\int dx \left[
        u \nabla f_1(t) - f_0(t) \nabla u 
    \right] \\ = - \int dx \left[
        f_1(t) + f_0(t) 
    \right] \nabla u = \int dx\, \sigma_\act \nabla u,
\end{multline}
where $\sigma_\act = - f_0 - f_1$ is the one-dimensional active stress.

In the framework of Hookean elasticity, the active stress can be introduced so as to shift the minimum of the potential \cref{eq:hooke_ela} by adding a symmetric tensor of \textit{active strain} $\bm{\epsilon}_\act$:
\begin{equation}\label{eq:hooke_t_act}
    U_t(\nabla\bm{u}) = \frac{\bar\lambda_t}{2} \big(\tr(\bm{\epsilon} + \bm{\epsilon}_\act)\big)^2 + \bar\mu_t \fr{\bm{\epsilon} + \bm{\epsilon}_\act}^2.
\end{equation}
From \cref{eq:tr,eq:fr} it follows then
\begin{align}\label{eq:tr_act}
    \big(\tr(\bm{\epsilon} + \bm{\epsilon}_\act)\big)^2
    =& (\tr\bm{\epsilon})^2 + 2 \tr(\bm{\epsilon}) \tr(\bm{\epsilon}_\act) + \cancel{\text{o.t.}}
    \nonumber\\=& (\tr\bm{\epsilon})^2 + 2 \nabla \cdot \bm{u} \tr(\bm{\epsilon}_\act),
\\\label{eq:fr_act}
    \fr{\bm{\epsilon} + \bm{\epsilon}_\act}^2
        =& \fr{\bm{\epsilon}}^2 + 2 \bm{\epsilon}_\act : \bm{\epsilon} + \cancel{\text{o.t.}}
        \nonumber\\=& \fr{\bm{\epsilon}}^2 + 2 \bm{\epsilon}_\act : \nabla\bm{u},
\end{align}
in which the constant offset terms are ignored as in \cref{eq:pot_t_osc}, since they do not affect the equations of motion. Using the identity
$$
    \tr(\bm{\epsilon}_\act) \mathds{1} : \nabla\bm{u}
        = \tr(\bm{\epsilon}_\act) \tr(\nabla\bm{u})
        = \tr(\bm{\epsilon}_\act) \nabla\cdot\bm{u},
$$
from \cref{eq:hooke_t_act,eq:tr_act,eq:fr_act} we obtain \cref{eq:pot_act} with
$$
    \bm{\sigma}_\act = \bar\lambda_t \tr(\bm{\epsilon}_\act)\mathds{1}
        + 2 \bar\mu_t \bm{\epsilon_\act},
$$
where $\bm{\epsilon}_\act$ is a given, explicit time-dependent or constant matrix.

\subsection{\label{app:fdt_chain}Rayleigh potential for harmonic chain}
To summarize the derivation of Rayleigh potential for linear elastodynamics, we start with the one-dimensional chain from \cref{app:lim}, following closely the presentation of Ref.~\cite[Chapter 15]{Tuckerman2023}. To this end it is convenient to recast the Lagrangian and the Newtonian equations of motion in the matrix-vector form denoted by the sans font. The Lagrangian~\cref{eq:lag_osci} and the respective Hamiltonian for the vector variable $\mtx{q} = \tra{(\delta X_1, \delta X_2,...,\delta X_N)}$ become
\begin{equation}\label{eq:lag_ham_chain}
    \mathcal{L} = \frac{1}{2} \left(\tra{\dot{\mtx{q}}}\mtx{M}\dot{\mtx{q}} - \tra{\mtx{q}}\mtx{K}\mtx{q}\right),\;
    \mathcal{H} = \frac{1}{2} \left(\tra{\dot{\mtx{q}}}\mtx{M}\dot{\mtx{q}} + \tra{\mtx{q}}\mtx{K}\mtx{q}\right),
\end{equation}
where $\mtx{M} = \operatorname{diag}(M_1, M_2,..., M_N)$ is a diagonal matrix of masses and
$$
    \mtx{K}_{ij} = \mtx{K}_{ji} = \begin{cases}
        \text{$K_{i-1} + K_i$ if $i = j$},\\
        \text{$-K_i$, if $j - i = 1$},\\
        \text{$-K_{i-1}$, if $i - j = 1$}\\
        \text{$0$ otherwise}.
    \end{cases}
$$
The Newtonian equations of motion corresponding to \cref{eq:lag_ham_chain} read
\begin{equation}\label{eq:mtx_chain}
    \ddot{\mtx{q}} = - \mtx{\Omega}^2 \mtx{q},
\end{equation}
where $\mtx\Omega^2 = \mtx{M}^{-1} \mtx{K}$, whose eigenvalues yield squared frequencies of the normal modes.

Generalizing the structure of \cref{eq:split_osc}, we select a part of the chain $i=1,2,...n$ as the subsystem of interest $\mtx{q}_\sys = \bm{q}_{1:n}$, while $\mtx{q}_\bath = \bm{q}_{n+1:N}$ represent the bath. Accordingly we can split \cref{eq:mtx_chain} into two parts
\begin{align}
    \label{eq:sys_chain}
    \ddot{\mtx{q}}_\sys =& - \mtx{\Omega}_{\sys\sys}^2 \mtx{q}_\sys - \mtx{\Omega}_{\sys\bath}^2 \mtx{q}_{\bath},
    \\\label{eq:bath_chain}
    \ddot{\mtx{q}}_\bath =& - \mtx{\Omega}_{\bath\bath}^2 \mtx{q}_\bath - \mtx{\Omega}_{\bath\sys}^2 \mtx{q}_\sys,
\end{align}
where the matrix $\mtx{\Omega}^2$ is divided into four blocks with coupling coefficients that correspond to elements describing the system, bath, and the couplings between them. Then we can treat the second term in \cref{eq:bath_chain} as a source and write the formal solution by using the Green function $\mtx{G}_{bb}(t) = \mtx{\Omega}_{\bath\bath}^{-1} \sin(\mtx{\Omega}_{\bath\bath} t)$ for $t\ge0$:
\begin{multline}\label{eq:qbath}
    \mtx{q}_\bath = \dot{\mtx{G}}_{\bath\bath}(t) \mtx{q}_\bath(0)
        + \mtx{G}_{\bath\bath}(t) \dot{\mtx{q}}_{\bath}(0)\\
        - \int_0^t dt' \mtx{G}_{\bath\bath}(t - t') \mtx{\Omega}_{bs}^2 \mtx{q}_{\sys}(t'),
\end{multline}
where $\dot{\mtx{G}}_{\bath\bath}(t) = \cos(\mtx\Omega_{\bath\bath} t)$
By substituting \cref{eq:qbath} into \cref{eq:sys_chain} and multiplying by $\mtx{M}_{\sys\sys}$, we obtain the equations of motion for the subsystem of interest $\mtx{q}_\sys$:
\begin{multline}\label{eq:qsys}
    \mtx{M}_{\sys\sys} \ddot{\mtx{q}}_\sys(t) = - \mtx{K}_{\sys\sys} \mtx{q}_\sys(t)
    + \mtx{K}_{\sys\bath}\int_0^t dt'\ \mtx{G}_{\bath\bath}(t - t') \mtx{\Omega}_{\bath\sys}^2 \mtx{q}_\sys(t')
    \\- \mtx{K}_{\sys\bath}\left[
        \dot{\mtx{G}}_{\bath\bath}(t) \mtx{q}_\bath(0)
        + \mtx{G}_{\bath\bath}(t) \dot{\mtx{q}}_{\bath}(0)
    \right].
\end{multline}

The bracketed term in \cref{eq:qsys} is entirely determined by the initial-value conditions of the bath $\big(\mtx{q}_\bath(0), \dot{\mtx{q}}_\bath(0)\big)$. In the equilibrium bath they are distributed canonically, cf. \cref{eq:boltzmann}: they are Gaussian with zero mean, $\mtx{q}_\bath(0)$ and $\dot{\mtx{q}}_\bath(0)$ are statistically independent with the covariance matrices $\kB T \mtx{K}_{\bath\bath}^{-1}$ and $\kB T \mtx{M}^{-1}$, respectively.

Therefore, as $N$ tends to infinity, the bracketed term in \cref{eq:qsys} turns into a spectral representation of a vector stochastic process~\cite[Chapter 2]{Ghanem2012}---the force generating spontaneous fluctuations
\begin{equation}\label{eq:flu_chain}
    \mtx{F}_\flu(t) =
        -\mtx{K}_{\sys\bath} \Big(
            \cos(\mtx\Omega_{\bath\bath} t) \mtx{q}_\bath(0)
        + \sin(\mtx\Omega_{\bath\bath} t) \mtx\Omega_{\bath\bath}^{-1} \dot{\mtx{q}}_\bath(0)
        \Big)
\end{equation}
with Fourier-like basis functions $\cos(\mtx\Omega_{\bath\bath} t)$, $\sin(\mtx\Omega_{\bath\bath} t)$, and random variates $\mtx{q}_\bath(0)$ and $\dot{\mtx{q}}_\bath(0)$ as coefficients. Indeed, if we diagonalize $\mtx{\Omega}_{\bath\bath} = \mtx{O}_\bath \mtx\Omega_\bath\mtx{O}_\bath^{-1}$ where the columns of orthogonal matrix $\mtx{O}_\bath$ and the elements of diagonal matrix $\mtx\Omega_\bath$ contain, respectively, eigenvectors and eigenvalues, we obtain a representation of the bath variables as normal modes $\hat{\mtx{q}}_\bath = \mtx{O}_\bath^{-1} \mtx{q}_\bath$ and their frequencies $(\Omega_\bath)_{a=1,2,\dots}$. Then from \cref{eq:flu_chain} we get
\begin{equation}\label{eq:flu_chain_modes}
    \mtx{F}_\flu(t) =
        -\mtx{K}_{\sys\bath} \mtx{O}_\bath \Big(
            \cos(\mtx\Omega_\bath t) \hat{\mtx{q}}_\bath(0)
            + \sin(\mtx\Omega_\bath t)\mtx\Omega_\bath^{-1}\dot{\hat{\mtx{q}}}_\bath(0)
        \Big),
\end{equation}
or in components
\begin{multline}\label{eq:fourier_chain}
    (\mtx{F}_\flu)_a = \sum_b \Big[
        c_{ab} \cos\big((\mtx{\Omega})_b t\big) + s_{ab} \sin\big((\mtx{\Omega})_b t\big)
    \Big]
    \\= \sum_b \Big[
        \frac{c_{ab} - i s_{ab}}{2} e^{i(\mtx{\Omega})_b t}
        + \frac{c_{ab} + i s_{ab}}{2} e^{-i(\mtx{\Omega})_b t}
    \Big]
\end{multline}
with Fourier-type expansion coefficients
\begin{align}
    c_{ab} =& -(\mtx{K}_{\sys\bath} \mtx{O}_\bath)_{ab} (\hat{\mtx{q}}_\bath)_b,\\
    s_{ab} =& -(\mtx{K}_{\sys\bath} \mtx{O}_\bath)_{ab} (\dot{\hat{\mtx{q}}}_\bath)_b.
\end{align}
Because these coefficients are derived as linear combinations of random variates 
$\big(\mtx{q}_\bath(0), \dot{\mtx{q}}_\bath(0)\big)$, the force $\mtx{F}_\flu(t)$ is also Gaussian characterized by the zero mean $\avg{\mtx{F}_\flu(t)} = \mtx{0}$. In the continuum limit \cref{eq:fourier_chain}, as $(c_{ab} - i s_{ab})/2\to d\Omega \hat{f}_\flu(\Omega)$, turns into the \textit{Karhunen-Loeve} expansion~\cite[Chapter 2]{Ghanem2012} of a Gaussian noise $f_\flu$ that takes the form of a Fourier transform
\begin{equation}
    f_\flu = \int_{-\infty}^{\infty} d\Omega\ \hat{f}_\flu(\Omega) e^{i\Omega t},
\end{equation}
in general with a nontrivial time autocorrelation function.

Furthermore, we can prove the fluctuation-dissipation theorem
\begin{equation}\label{eq:fdt}
    \avg{\mtx{F}_\flu(0) \tra{\mtx{F}_\flu}(t)} = \kB T \mtx{\Gamma}(t),
\end{equation}
with a \textit{dissipative memory kernel} $\mtx\Gamma(t)$, by calculating the covariance matrix directly from \cref{eq:flu_chain}:
$$
    \mtx{\Gamma}(t) = \beta \avg{\mtx{F}_\flu(0) \tra{\mtx{F}_\flu}(t)} = \mtx{K}_{\sys\bath} \mtx{K}_{\bath\bath}^{-1} \cos(\mtx{\Omega}_{\bath\bath} t) \mtx{K}_{\bath\sys},
$$
in which we have used the statistical independence of $\mtx{q}(0)$ and $\dot{\mtx{q}}(0)$ together with the symmetry of $\mtx{K}$ and $\mtx{\Omega}_{\bath\bath}$. We also observe
\begin{multline}\label{eq:mem_chain}
    \dot{\mtx\Gamma}(t) = -\mtx{K}_{\sys\bath} \mtx{K}_{\bath\bath}^{-1} \mtx{\Omega}_{\bath\bath} \sin(\mtx{\Omega}_{\bath\bath} t) \mtx{K}_{\bath\sys}
    \\= -\mtx{K}_{\sys\bath} \mtx{\Omega}_{\bath\bath}^{-1} \sin(\mtx{\Omega}_{\bath\bath} t) \mtx{M}_{\sys\sys}^{-1} \mtx{K}_{\bath\sys}
    = - \mtx{K}_{\sys\bath} \mtx{G}_{\bath\bath}(t) \mtx{\Omega}_{\bath\sys}^2.
\end{multline}

Note that the matrices $\mtx{K}_{\sys\bath} = \tra{\mtx{K}_{\bath\sys}}$ contain only two nonzero elements, that couple the bath to the subsystem's ends at $(\mtx{q}_\sys)_1$ and $(\mtx{q}_\sys)_n$. Other components of $(\mtx{F}_\flu)_{i\not\in\{1,n\}}$ vanish identically. Being separated by infinitely many oscillators, in the limit $N\to\infty$, they can be assumed statistically independent. 

With help of \cref{eq:flu_chain,eq:mem_chain}, we can now rewrite \cref{eq:qsys} as a generalized Langevin equation
\begin{equation}\label{eq:gle_chain}
    \mtx{M}_{\sys\sys} \ddot{\mtx{q}}_\sys(t) + \mtx{K}_{\sys\sys} \mtx{q}_\sys(t)
    = - \int_0^t dt'\ \dot{\mtx{\Gamma}}(t - t') \mtx{q}_\sys(t') + \mtx{F}_\flu.
\end{equation}
A more common---though less general---form of \cref{eq:gle_chain}, which involves convolution of the memory kernel with the velocities, can be obtained through integration of the first term by parts~\cite[Chapter 15]{Tuckerman2023}:
\begin{multline}
    \label{eq:gle_chain_vel}
    \mtx{M}_{\sys\sys} \ddot{\mtx{q}}_\sys(t) + \mtx{K}_{\sys\sys} \mtx{q}_\sys(t)
    = \mtx{\Gamma}(t) \mtx{q}_\sys(0) - \mtx{\Gamma}(0) \mtx{q}_\sys(t) \\- \int_0^t dt'\ \mtx{\Gamma} \dot{\mtx{q}}_\sys(t') + \mtx{F}_\flu.
\end{multline}
In a special case where the covariance between $\mtx{q}_\sys(0)$ and $\mtx{q}_\bath(0)$ vanishes, like the one considered in Ref.~\cite[Chapter 15]{Tuckerman2023}, the term $\mtx\Gamma(t) q_\sys(0)$ can be incorporated into the stochastic force $\mtx{F}_\flu$ without consequences for the fluctuation-dissipation theorem \cref{eq:fdt}. Here, due to the structure of stiffness matrix $\mtx{K}$, this term would introduce a $\mtx{K}_{\sys\bath}$-dependent factor into~\cref{eq:fdt}.

Equation~\cref{eq:gle_chain} is fully consistent with the prototypical Rayleigh function~\cref{eq:ray_proto} in the limit of $\Gamma(t - t') \to \gamma\delta(t - t')$, when the components of $(\mtx{F}_\flu)_{i\in\{1,n\}}$ become statistically independent. Indeed, from Eq.~\eqref{eq:gle_chain} we then get
\begin{equation}\label{eq:langevin_chain}
    M_{\sys\sys} \ddot{\mtx{q}}_\sys(t) + \mtx{K}_{\sys\sys} \mtx{q}_\sys(t)
    = - \mtx{\gamma} \dot{\mtx{q}}_\sys(t) + (2\kB T \mtx\gamma)^{1/2} \dot{\mtx{w}}(t),
\end{equation}
where $(\mtx{w})_{i\in\{1,n\}}$ are standard independent Wiener process and $(\mtx{w})_{i\not\in\{1,n\}} \equiv 0$.

Finally, while the left-hand side of \cref{eq:gle_chain} can be derived from the Lagrangian
$$
    \mathcal{L}_\sys = \frac{1}{2}\left(
        \dot{\mtx{q}}_\sys^T \mtx{M}_{\sys\sys} \dot{\mtx{q}}_\sys - \mtx{q}^T_\sys \mtx{K}_\sys \mtx{q}_\sys
    \right)
$$
the right hand side suggests a definition of the Rayleighian
\begin{equation}\label{eq:ray_memo_chain}
    \rayleigh = \tra{\dot{\mtx{q}}_\sys}\left(\int_0^t dt'\ \dot{\mtx{\Gamma}} \mtx{q}_\sys - \mtx{F}_\flu\right),
\end{equation}
or, in the white-noise limit of \cref{eq:langevin_chain},
\begin{equation}\label{eq:ray_delta_chain}
    \rayleigh = \tra{\dot{\mtx{q}}_\sys}\left(\frac{\mtx\gamma}{2}\dot{\mtx{q}}_\sys - (2\kB T \mtx\gamma)^{1/2}\dot{\mtx{w}}_\flu\right).
\end{equation}

\subsection{\label{app:fdt_hooke}Rayleigh potential for a Hookean solid}
The theoretical methods of \cref{app:fdt_chain} also apply to the Hookean model of linear elasticity and lead to the Rayleigh potential with a dissipative kernel. In a complete analogy to the Newtonian \cref{eq:mtx_chain}, we consider Cauchy \cref{eq:cauchy}, extended beyond the body region $\xi \in (0, \infty)$ to describe the environment medium as a an elastic material, in the form
\begin{equation}\label{eq:cauchy_wave}
    \mathcal{W}\bm{u} = (\partial_t^2 + \mathcal{D}^2) \bm{u} = 0,
\end{equation}
in which
$$
    \mathcal{D}^2 \bm{u} = -\bar{\rho}^{-1}\nabla \cdot \Big(
        \lambda \bar{\bm{g}} \tr{\nabla\bm{u}} + \mu (\nabla \bm{u} + \nabla \bm{u}^T)
    \Big).
$$
We split the Cauchy equation of motion over the system's and bath's domains $\Xi_\sys = [0, 1]$ and $\Xi_\bath = (1, \infty)$, respectively, 
\begin{align}\label{eq:sys_wave}
    \mathcal{W} \bm{u}_\sys =& - \frac{1}{\bar{\rho}}\frac{\delta(\xi - 1)}{\sqrt{\bar{g}}} \bar{\bm{n}} \cdot \bm{\sigma}_\bath,
    \\\label{eq:bath_wave}
   \mathcal{W} \bm{u}_\bath =& - \frac{1}{\bar{\rho}} \frac{\delta(\xi - 1)}{\sqrt{\bar{g}}} \bar{\bm{n}} \cdot \bm{\sigma}_\sys,
\end{align}
which parallel \cref{eq:sys_chain,eq:bath_chain} with the source term \cref{eq:dirac_delta} arising due the new boundary condition
$$
    \bm{\sigma}_\sys|_{\xi=1} = \bm{\sigma}_\ela|_{\xi=1} = \bm{\sigma}_\bath|_{\xi=1}. 
$$

A complete analogy between \cref{eq:sys_chain,eq:bath_chain} and \cref{eq:sys_wave,eq:bath_wave} can be established if the latter is converted to the matrix notation in normal modes. Given proper boundary conditions, the operator $\mathcal{D}^2$, derived from a quadratic potential functional $U$, has eigenvector fields $\bm{\psi}_i$ with eigenvalues $\kappa_i$, which form an orthonormal basis set of functions $\vert \bm{\psi}_i \rangle$. Similarly we can obtain the sets of such eigensystems for the domains of $\Xi_\sys$ and $\Xi_\bath$, and write
$$
    \bm{u}_\sys = \sum_i (\mtx{u}_\sys)_{i} \vert (\bm{\psi}_\sys)_i \rangle,\quad
    \bm{u}_\bath = \sum_i (\mtx{u}_\bath)_i \vert (\bm{\psi}_\bath)_i \rangle,
$$
where $(u_\any)_i$ are Fourier coefficients with $\any=\sys,\bath$. Then we can introduce the density and frequency matrices:
$$
    (\mtx\varrho_\any)_{ij} = \langle (\bm{\psi}_\any)_i \vert \bar\rho \vert (\bm{\psi}_\any)_j \rangle,\quad
    \mtx\Omega = (\mtx\varrho_\any^{-1} \mtx{K}_\any)^{1/2},\quad
$$
where $(\mtx{K}_\any)_{ij} = (\kappa_\any)_i \delta_{ij}$, and the coupling coefficients
\begin{align}
    (\mtx{C}_{\bath\sys})_i =& \left\langle (\bm{\psi}_\bath)_i \bigg|
        \frac{1}{\bar{\rho}} \frac{\delta(\xi - 1)}{\sqrt{\bar{g}}} \bar{\bm{n}} \cdot \bm{\sigma}_\ela
    \right\rangle,\\
    (\mtx{C}_{\sys\bath})_i =& \left\langle (\bm{\psi}_\sys)_i \bigg|
        \frac{1}{\bar{\rho}} \frac{\delta(\xi - 1)}{\sqrt{\bar{g}}} \bar{\bm{n}} \cdot \bm{\sigma}_\ela
    \right\rangle.
\end{align}
Now we can recast \cref{eq:sys_wave,eq:bath_wave} as
\begin{align}\label{eq:sys_wave_mtx}
    (\ddot{\mtx{u}}_\sys)_i =& - (\kappa_\sys)_i\ (\mtx{u}_\sys)_i - (\mtx{C}_{\sys\bath})_i,
    \\\label{eq:bath_wave_mtx}
    (\ddot{\mtx{u}}_\bath)_i =& - (\kappa_\bath)_i\ (\mtx{u}_\bath)_i - (\mtx{C}_{\bath\sys})_i.
\end{align}

The last two equations are analogous to the diagonalized version of \cref{eq:sys_chain,eq:bath_chain} in terms of normal modes. Therefore the analysis of \cref{app:fdt_chain} that follows after \cref{eq:bath_chain} is applicable to the continuum case. It only remains to interpret the results expressed in normal modes for differential operators over vector fields $\bm{u}$. The memory kernel $\mtx\Gamma$ and the stochastic force $\mtx{F}_\flu$ correspond to a time-dependent operator $\mathcal{G}(t)$ and a random force $\bm{f}_\flu = \bm{\nabla}\cdot\bm{\sigma}_\flu$ derived from a fluctuating stress $\bm{\sigma}_\flu$.

To obtain \cref{eq:ray_delta_wave}, we assume the Rayleighian
\begin{equation}\label{eq:ray_memo_wave}
    \rayleigh = \int_{\bar{V}} d\bar{V}\ \partial_t\bm{u} \cdot \left(\int_0^t dt'\ \mathcal{G}(t -t') \bm{u}(t') - \bm{f}_\flu\right),
\end{equation}
with a delta-correlated short-range memory operator $$
    \mathcal{G}(t - t')\bm{u} = \bar{\gamma} \frac{\bar{n}}{\sqrt{\bar{g}}}\delta(\xi-1) \dot\delta(t - t') \bm{u}(t')
$$ fully localized to the surface $\partial\bar{V}$.

The \citet{LandauLifshitz1959Fluctuations} model of fluctuating hydrodynamics can also be cast in the form of \cref{eq:ray_memo_wave} if we consider a longer-range memory operator $\mathcal{G}$. To this end it is convenient to introduce a viscous Lam\'e coefficients $\tilde\lambda$ and $\tilde\mu$, and Lam\'e tensor
$$
    \Lambda^{\ell mjk}(\bar{g}) = \tilde\lambda \bar{g}^{\ell m} \bar{g}^{jk} + \tilde\mu (\bar{g}^{\ell j} \bar{g}^{m k} + \bar{g}^{\ell k} \bar{g}^{mj}).
$$
With their help we can define the memory operator
\begin{multline}\label{eq:hydro}
    (\mathcal{G}(t-t')\bm{u})^i = \dot\delta(t-t') \int_{\bar{V}} d\bar{V}'\ G^{ij}(\bar{\bm{B}}, \bar{\bm{B}}')\  u_j(\bar{\bm{B}}')
    \\=\dot\delta(t-t')\int_{\bar{V}} d\bar{V}' \nabla_j \left[
    \tilde\lambda \bar{g}^{ij} \nabla_k \delta^3
    \right.\\\left.
    + \tilde\mu \left(\delta^j_k \nabla^i \delta^3  + \delta^i_k \nabla^j\delta^3 \right)\right] u^k,  
\end{multline}
where
$$
    G{^{i j}} = \nabla_\ell [\Lambda{^{i \ell j m}} \nabla_m \delta^3(\bar{\bm{B}}, \bar{\bm{B}}')].
$$

\subsection{\label{app:particle}Particle motion}
Interestingly, in the macroscopic limit, when the shape variations of a Hookean elastic body are negligible, the Langevin equation for its center of mass can be derived directly from \cref{eq:most_general_wave}. In this regime, the density may be considered approximately uniform, $\bar{\rho} \approx \mass / \bar{V}$. Then the body's center of mass coincides with its geometric center
\begin{equation}\label{eq:com}
    \bm{\com} \approx \bar{V}^{-1} \int_{\bar{V}} d\bar{V}\ (\bar{\bm{B}} + \bm{u}) \approx \bar{S}^{-1} \int_{\partial\bar{V}}d\bar{S} (\bar{\bm{B}} + \bm{u})
\end{equation}
in which $\bar{S} = \int_{\partial\bar{V}} dS$. By integrating \cref{eq:most_general_wave} over $\bar{V}$, we obtain equations of motion
\begin{equation}\label{eq:com_eom}
    \mass \ddot{\bm{\com}} + \alpha \dot{\bm{\com}}  = \bm{F} + \sqrt{2D \alpha}\ \dot{\bm{w}},
\end{equation}
with the friction constant $\alpha = \bar\gamma\bar{S}$, the diffusion coefficient
\begin{equation}\label{eq:brownian} D = \kB T / \alpha, \end{equation}
and standard delta-correlated Gaussian white noise
\begin{equation}\label{eq:noise}
    \dot{\bm{w}} = \bar{S}^{1/2}\int_{\partial\bar{V}} dS\ \partial\bm{w}.
\end{equation}

The total stress $\bm{\sigma} = \bm{\sigma}_\ela + \bm{\sigma}_\act + \bm{\sigma}_\ext$ integrated over the body provides here the deterministic force
\begin{equation}\label{eq:force_particle}
    \bm{F} = \int_{\bar{V}} d\bar{V}\ \bm{\nabla}\cdot\bm{\sigma}
        = \int_{\partial \bar{V}} d\bar{\bm{S}} \cdot \bm{\sigma}.
\end{equation}
Moreover the conservative part of this force retains its relation to the potential gradient 
\begin{multline}\label{eq:force_U}
    \bm{F}_U = \int_{\bar{V}} d\bar{V}\ \bm{\nabla}\cdot(\bm{\sigma}_\ela + \bm{\sigma}_\act) \\= - \int_{\bar{V}}d\bar{V}\ \frac{\delta U_\act}{\delta\bm{u}} \cdot \left(\frac{\delta\bm{\com}}{\delta\bm{u}}\right)^{-1} = -\frac{\partial U_\act}{\partial\bm{\com}},
\end{multline}
since from~\cref{eq:com} we have $\delta\bm{\com}/\delta\bm{u} = \mathds{1}$.

Note that the overdamped limit of \cref{eq:com_eom}
$$
    \alpha \dot{\bm{\com}}  = -\frac{\partial U_\act}{\partial\bm{\com}} + \bm{F}_\ext + \sqrt{2D \alpha}\ \dot{\bm{w}},
$$
can be treated as the equation of motion for the geometric center of the body without invoking the uniform-density approximation,---because \cref{eq:noise,eq:force_particle,eq:force_U} are independent of this assumption.

Remarkably, by using \cref{eq:Qdef}, we can also examine the Sekimoto definition of heat~\cite{Sekimoto1998} in stochastic thermodynamics. The integral appearing in this equation can be approximated, if we neglect the fluctuations of $\partial_t\bm{u}$ over the surface of the body, thus assuming a translational rigid-body motion $\partial_t\bm{u} \simeq \dot{\bm{\com}}$. In such case
\begin{equation}\label{eq:sekimoto}
    \dot{Q} \approx \alpha \dot{\bm{\com}} \cdot (\dot{\bm{\com}} - \sqrt{2 D \alpha} \dot{\bm{w}}),
\end{equation}
coincides with the Sekimoto definition of heat.

\subsection{\label{app:mic}Hydrodynamics at microscopic scales}
To derive the microscopic equation of motion \eqref{eq:mic}, we take the time derivative of the momentum density $\bm\pi = \rho \bm{v}$ conjugate to the barycentric coordinate $\bm{u}(\bar{\bm{B}})$:
\begin{multline}\label{eq:time_derivative}
    \partial_t \pi^a = \int_\cell \frac{d\cell(\bm{r})}{\cell} \\\times \sum_i \left[
        \ddot{q}^a(i)
        - \dot{q}^a(i) \dot{q}^b(i) \nabla_b
    \right] \rho_i\big(\bar{\bm{B}} + \bm{r} - \bm{q}(i)\big),
\end{multline}
in which the operator inside the bracket contains two contributions. In the barycentric coordinates $\bm{r}(i) = \bm{q}(i) - \bm{C}$ we substitute the \textit{peculiar} velocities $\dot{\bm{r}}(i)$ for $\dot{\bm{q}}(i) = \bm{v} + \dot{\bm{r}}(i)$ to split the kinetic part
\begin{multline}\label{eq:kinetic}
    \int_\cell \frac{d\cell(\bm{r})}{\cell} \sum_i \dot{q}^a(i) \dot{q}^b(i) \nabla_b
    \rho_i
    \\= \bar\nabla_b \sum_i \dot{r}^a(i) \dot{r}^b(i)
    \rho_i\big(\bar{\bm{B}} + \bm{r} - \bm{q}(i)\big)
    \\+ \rho v^b(\bar{\bm{B}}) \bar\nabla_b v^a(\bar{\bm{B}}),
\end{multline}
where we invoke the identities $\sum_i \dot{\bm{r}}(i) \equiv \bm{0}$ and$$
    \nabla_b \rho_i\big(\bar{\bm{B}} + \bm{r} - \bm{q}(i)\big) = \frac{\partial \rho_i}{\partial r^b} = \frac{\partial \rho_i}{\partial \bar{B}^b} = \bar\nabla_b \rho_i\big(\bar{\bm{B}} + \bm{r} - \bm{q}(i)\big).
$$
Thereby in \cref{eq:kinetic} we separate the kinetic contribution of the stress from the convective term, which is expected in the Lagrangian reference frame, since the unit cell position $\bar{\bm{B}}$ is not moving with the center of mass.

The potential part of the operator inside the bracket in \cref{eq:time_derivative} yields
\begin{multline}
    \int_\cell \frac{d\cell(\bm{r})}{\cell} \sum_i \ddot{q}^a(i) \rho_i
    \\= \int_\cell \frac{d\cell(\bm{r})}{m \cell} \sum_i \frac{U_\mic}{\partial q_a(i)} q^b \nabla_b\rho_i \big(\bar{\bm{B}} + \bm{r} - \bm{q}(i)\big)
    \\= \bar\nabla_b \int_\cell \frac{d\cell(\bm{r})}{\cell} \sum_i \frac{U_\mic}{\partial q_a(i)} q^b \delta\big(\bar{\bm{B}} + \bm{r} - \bm{q}(i)\big),
\end{multline}
which constitutes the potential term of \cref{eq:sigma}.

\subsection{\label{app:rot}Infinitesimal and finite rotations}
The linear theory of Hookean elasticity formulated for an infinite medium neglects quadratic terms in \cref{eq:g_u,eq:neighborhood}, thus rendering the Hamiltonian symmetric with respect to infinitesimal rotations described by the tensor $\bm\omega$ in \cref{eq:nabla_u}. When the same linearized theory is applied to a bounded object, this symmetry introduces a modeling artifact under finite displacements, because the elastic energy remains invariant under affine transformations of the form $\bm{1}+\bm\omega$. While indistinguishable to linear order in $\bm\omega$ from a rigid rotation
\begin{equation}\label{eq:rotation}
    \bm{\mathsf{R}} = e^{\bm\omega} = \bm{1} + \bm\omega + O(\bm\omega^2),
\end{equation}
such a deformation is not an exact rotation: at second order it represents a physical stretch resulting in an increase of the occupied volume.

Whereas changes of the system volume due to small perturbations $\bm\omega$ are ignored in questions involving energy density of an infinite medium, a finite elastic body is usually expected to be symmetric under finite rotations $\bm{\mathsf{R}}$ rather than under the transformations $\bm{1}+\bm\omega$. This modeling artifact therefore must be properly handled in numerical simulations. By integrating equations of motion over time $t_i=i \Delta t$ with a step $\Delta t$ ($i=0,1,....)$, we decompose the displacements into the body's deformation $\bm{u}(t,\bm\Omega)$ in the co-moving coordinate frame, alongside the rotation $\bm{\mathsf{R}}(t)$ and translation $\bm A(t)$ with respect to the centroid of its reference configuration in the \textit{laboratory} reference frame. We project out the global translational and infinitesimal rotational modes of displacement at each integration step, and accumulate the extracted increments as the total translation $\bm{A}(t)$ and a finite orientation
\begin{equation}
    \bm{\mathsf{R}}(t_i + \Delta t) = \bm{\mathsf{R}}(t_i) \cdot \exp(\Delta\bm\omega),
\end{equation}
of the body relative to the chosen \textit{laboratory} coordinates. Here $\Delta\bm\omega$ contains the asymmetric changes of the deformation gradient due to $\bm{f}_\ext$ and $\bm{f}_\flu$~\SM{Sec.~\smNum}. To simulate external forces acting on the body from the laboratory reference frame, we correspondingly update the material coordinates of the surface point $(\theta_\ext, \phi_\ext)$ where the force is applied, as the body is rotated through $\bm{\mathsf{R}}(t)$~\SM{Sec.~\smNum}.

\bibliography{refs}

\end{document}